\documentclass[aps,preprint,showpacs,preprintnumbers,amsmath,amssymb,
floatfix] {revtex4}
\usepackage{graphicx}
\usepackage[all]{xy}
\usepackage{dcolumn} 
\usepackage{longtable}
\usepackage{lscape}

\begin{document}
\title{Skyrme Interaction and Nuclear Matter Constraints} 
\author{M. Dutra\footnote{Present address: Departamento de F\'isica, Instituto 
Tecnol\'ogico da Aeron\'autica, CTA, S\~ao Jos\'e dos Campos, 12228-900. SP,
Brazil},
O. Louren\c co\footnote{Present address: Departamento de F\'isica, Instituto
Tecnol\'ogico da Aeron\'autica, CTA, S\~ao Jos\'e dos Campos, 12228-900. SP,
Brazil},
J. S. S\'a Martins, and A. Delfino}
\affiliation{Instituto de F\'{\i}sica - Universidade Federal Fluminense,
Av. Litor\^ anea s/n, 24210-150 Boa Viagem, Niter\'oi RJ, Brazil}
\author{J. R. Stone}
\affiliation{Oxford Physics, University of Oxford, OX1 3PU Oxford, 
United Kingdom}
\affiliation{Department of Physics and Astronomy, University of Tennessee, 
Knoxville, Tennessee 37996, USA}
\author{P. D. Stevenson}
\affiliation{Department of Physics, University of Surrey, Guildford, GU2 7XH UK}
\date{\today}

\begin{abstract}
This paper presents a detailed assessment of the ability of the 240 Skyrme 
interaction parameter sets in the literature to satisfy a series of criteria
derived from macroscopic properties of nuclear matter in the vicinity of nuclear
saturation density at zero temperature and their density dependence,
derived by the liquid drop model, experiments with giant resonances and
heavy-ion collisions. The objective is to identify those parameterizations which
best satisfy the current understanding of the physics of nuclear matter over a
wide range of applications. Out of the 240 models, only 16 are shown to satisfy
all these constraints. Additional, more microscopic, constraints on density
dependence of the neutron and proton effective mass beta-equilibrium matter,
Landau parameters of symmetric and pure neutron nuclear matter, and
observational data on high- and low-mass cold neutron stars further reduce this
number to 5, a very small group of recommended Skyrme parameterizations to be
used in future applications of the Skyrme interaction of nuclear matter related
observables. Full information on partial fulfillment of individual constraints
by all Skyrme models considered is given. The results are discussed in terms of
the physical interpretation of the Skyrme interaction and the validity of its
use in mean-field models. Future work on application of the Skyrme forces,
selected on the basis of variables of nuclear matter, in Hartree-Fock
calculation of properties of finite nuclei, is outlined.
\end{abstract}

\maketitle

\section{\label{sec:intro}Introduction}

Empirical properties of infinite nuclear matter can be calculated using many 
different theoretical approaches. The most microscopic ones start from a
realistic two-body free nucleon-nucleon (NN) interaction with parameters fitted
to NN scattering phaseshifts in different partial wave channels and to
properties of the deuteron \cite{modern}. Taking these bare NN interactions as
input into many-body formalism, such as the relativistic
Dirac-Bruckner-Hartree-Fock (DBHF) approximation and its
non-relativistic counterpart BHF \cite{dbhf,bhf}, variational methods 
\cite{akmal1998,mukherjee2007}, Correlated Basis Function \cite{cbf}, 
Self-Consistent Greens Function models (SCGF) \cite{dewulf2003, frick2003},
Quantum Monte Carlo techniques \cite{pudinger1997, schmidt1999, carlson2003,
gandolfi2007, gandolfi2008, gandolfi2009} and Chiral Effective Field Theory
\cite{hebeler2010,hebeler2010a} an effective NN interaction, which includes the
effect of the medium, is derived and the many-body problem approximately
solved. 

Various many-body approaches typically lead to an over prediction of the
saturation density $\rho_{\rm o}$ of symmetric nuclear matter (SNM), at which
the binding energy per nucleon reaches its maximum, and of the corresponding
maximum binding energy $E_{\rm o}$($\rho$=$\rho_{\rm o}$) \cite{li2006}. There
are many ways of estimating the experimental value of $\rho_{\rm o}$, including
different variants of the liquid drop models, optical model of NN scattering,
muonic atoms, and Hartree-Fock (HF) calculation of nuclear density distributions
(see e.g. \cite{haensel1981} and references therein). The range of results is
rather broad but a consensus value is $\rho_{\rm o}$=0.17$\pm$0.03 fm$^{\rm
-3}$.  The empirical value $E_{\rm o}$ per nucleon $\sim$ 16 MeV can be
extracted from the semi-empirical mass formula or from the extrapolation of
binding energies of heavy nuclei. Theoretical calculation of saturation
properties of SNM is not only dependent on the choice of the bare NN interaction
but also on the method of treatment of  many-body effects. For
example, if the BHF  approximation is used, $E_{\rm o}$ and $\rho_{\rm o}$ are
correlated within a narrow band \cite{li2006,coester1970}. Two main approaches
have been suggested to improve the theoretical calculation of saturation
properties of SNM, the most frequently used being the inclusion of 3-body (NNN)
forces. As the form of these forces is unknown, different \textit{ad hoc}
parameterizations have been used, dependent on additional variable parameters
that need to be fitted to account for the delicate balance between the strong
(NNN) attraction and (NN) repulsion at short distances. Alternatively,  DBHF
calculations have been shown to be effective without the need for NNN forces
\cite{li2006,li1992}. Another possibility is to treat the scalar and vector
densities in the Walecka relativistic mean-field model \cite{walecka1986} as
equal \cite{delfino2006}. However, the systematic deviation of all theoretical
predictions from the expected empirical values of $E_{\rm o}$ and $\rho_{\rm o}$
remains a problem. An interesting suggestion by Dewulf {\it et al.}
\cite{dewulf2003} implies that treatment of short-range correlations in nuclear
matter in the SCGF model brings the saturation density closer to the empirical
value than do current BHF calculations. Careful examination of the effect of
long-range correlations on nuclear saturation properties through coherent
pion-exchange contributions to the binding energy of nuclear matter is equally
important. Such correlations are not present in finite nuclei and a question
arises as to how applicable are liquid drop model predictions, based on
properties of finite nuclei, to saturation properties of nuclear matter.

Quite a different perspective on treating many-nucleon systems is to use 
effective density dependent NN and NNN interactions instead of realistic ones.
The pioneering models of K\"{o}hler \cite{kohler1965}, Brink and Boeker
\cite{brink1967}, Moszkowski \cite{mosz1970},  Skyrme \cite{skyrme1965}, further
developed by Vautherin and Brink (see \cite{vautherin1972} and references
therein) and Gogny \cite{decharge1980} initiated this approach, widely used
today. The basic idea is to parameterize the NN and NNN  interactions by zero
range (Skyrme model), short finite range (Gogny model) and indefinite range
(Separable Monopole model (SMO) \cite{stevenson2001,stone2002}) density
dependent functionals to describe the ground state properties of finite nuclei
and nuclear matter. In this scenario, the microscopic details of NN and NNN
forces, such as meson exchange, are not explicitly considered and all the
physically relevant information is carried by the parameters  of the density
dependent phenomenological forces which include the spin, orbital angular
momentum and isospin couplings. The drawback of this approach is that the
parameterization of such forces is not unique and there exist, in principle, an
infinite number of parameter sets, fitted to ground state properties of (doubly-
or semi-magic) stable nuclei, fission barriers, energies of giant resonances and
symmetric and asymmetric nuclear matter (ANM). This situation arises in part
because there is no unambiguous connection between individual parameters, or
groups of parameters, of these forces with particular physical properties of the
many-body nuclear system. Many parameters are strongly correlated.

It is obviously desirable to constrain the parameterizations of effective
density dependent forces as much as possible. The strategy chosen in this work
is to concentrate first on application of the forces to modelling different
variants of infinite nuclear matter. Although nuclear matter is an idealized
medium, and all its properties, derived from experiments indirectly in a  model
dependent way, are \textit{empirical} quantities, it offers an important insight
into specific parts of the phenomenological interaction and has important
applications in the theory of heavy-ion collisions (HIC) and the physics of neutron
stars. This paper focuses on the Skyrme interaction and is the first part of a
series in which nuclear matter constraints will be applied to each class of
effective density dependent interaction. It will be followed by investigation
of Gogny and SMO forces and interactions used in relativistic mean-field (RMF)
models. After this work is complete, the implications of the consequences for
the theory of finite nuclei will be investigated. 

We use the most up-to-date constraints  on nuclear matter properties which go
much beyond the minimal conditions on SNM (saturation density, binding energy,
incompressibility and symmetry energy at saturation). New data from HIC, giant 
monopole and dipole resonance experiments as well as new observational data on
neutron stars provide new constraints on the performance of individual Skyrme
parameterizations in nuclear matter. 

The set of eleven macroscopic constraints used in this work have been mainly 
derived from experimental data, on the assumption of the validity of the
liquid-drop model \cite{myers1969,myers1974}, and concern properties of SNM at
and close to the saturation point. Studies of dilute Fermi gas provide
constraints on low density pure neutron matter (PNM) equation of state (EoS). 
Dynamical models of HIC further constrain the density dependence of pressure in
SNM and PNM at subsaturation density and extrapolate these constraints to higher
densities. Mean-field Hartree-Fock + Random Phase Approximation (RPA), used in
calculating giant resonance excitation energies both in relativistic and
non-relativistic models, provide a final group of constraints on the
incompressibility of nuclear matter and its density and symmetry dependence.

In addition to the above macroscopic constraints, several more microscopic 
constraints are employed. These include the density dependence of the nucleon 
effective masses  in beta-equilibrium matter (BEM), the Landau  parameters for
SNM and PNM and of the symmetry energy. Observational data on cold
non-rotational high-mass and low-mass neutron stars provide a final group of
constraints.
 
In this work we consider 240 Skyrme parameter sets, currently available in the 
literature, and critically compare their predictions for a wide variety of
properties of SNM, PNM, asymmetric matter with fixed ratio proton fraction (ANM)
and BEM with all available constraints  in the density
range from $\sim$0.1$\rho_{\rm o}$ to 3$\rho_{\rm o}$, estimated on the best
available experimental and theoretical grounds. The range of applicability of
the Skyrme force is a very important issue which is often mishandled. 

The Skyrme interaction, originally constructed for finite nuclei and nuclear 
matter at saturation density, is a low momentum expansion of the effective
two-body NN interaction in momentum space, and both the lower and upper limits
of its validity are not firmly established. The important point about the Skyrme
interaction is that some correlation effects are included through its
parameters. Thus, although formulated as zero-range in coordinate space
\cite{vautherin1972}, it exhibits some finite-range features \cite{pethick1995}.

For finite nuclei, the best evidence for a lower limit derives from the fact 
that Skyrme models reasonably predict the observed abrupt decrease of
density at the nuclear surface, and neutron and proton mean-square
radii. The sensitivity extends down to about 0.1$\rho_{\rm o}$.

In uniform SNM the Skyrme interaction has been used to make prediction of the 
appearance of light clusters (deuterons, tritons, $^{\rm 3}$He and
alpha-particles in hot matter in the region of density of 0.6 - 1.25$\rho_{\rm
o}$ \cite{roepke1983}). More recently, abundance of light clusters with $A\le$13
in supernova envelopes at finite temperature was calculated at density range
0.01 - 0.5$\rho_{\rm o}$ using the Skyrme functional \cite{heckel2009}. The
``pasta'' phase, predicted in neutron star and supernova matter in a variety
of models \cite{pasta1,pasta2}, was successfully modeled in supernova matter in
a density range  0.25 - 0.75$\rho_{\rm o}$ in self-consistent HF+BCS calculation
with SkM* and SLy4 Skyrme interaction \cite{newton2009}. 

PNM has been mainly studied as an approximation to a low density Fermi gas. 
Schwenk and Pethick \cite{schwenk2005} explored, in a model independent way, the
neutron matter EoS at densities 0.0125 - 0.125$\rho_{\rm o}$, and Epelbaum {\it
et al.} \cite{epelbaum2009} calculated the ground state energy of dilute neutron
matter at next-to-leading-order in lattice chiral effective field theory in the
density range 0.02 - 0.1$\rho_{\rm o}$. Quantum Monte Carlo techniques have been
applied to low density PNM, providing a constraint  for on the EoS up to
saturation density \cite{gezerlis2008, gezerlis2010}. To our knowledge, there
has not yet been a detailed study of the applicability of the Skyrme interaction
at these low densities in PNM. Such a study is of a particular interest as it
may be one of the best ways to model the crust of neutron stars.  

The upper density limit of validity of the Skyrme interaction reflects the fact 
that at higher densities relativistic effects should be increasingly important.
The appearance of heavy strange baryons and mesons in the matter is ultimately
inevitable. Due to Pauli blocking, the chemical potential of the neutrons
increases rapidly with density. At some point, it becomes energetically
favourable for the system to let the neutrons undergo a strangeness changing
weak decay, which replaces them by hyperons, for which the Fermi sea is not yet
filled. From the difference of mass between the neutron and its strange partners
it follows that the critical density, at which hyperons should appear, is
2-3$\rho_{\rm o}$. Using a nucleon-only Skyrme interaction beyond this density
can be expected to yield misleading results. This is discussed later in
connection with high-mass neutron star models. 

Taking all the above pieces of evidence into account, we adopt 0.01$\rho_{\rm o}
 \le \rho \le$ 3$\rho_{\rm o}$ as the range of validity of the Skyrme
interactions considered in this work. 

The paper is organised as follows. A brief description of the Skyrme
interaction, together with definition of the variables used in this work, is
 given in Sec.~\ref{sec:skyrme}. Classification of the macroscopic constraints
and discussion of their origin and applicability range forms the content of
Sec.~\ref{sec:macro}. Sec.~\ref{sec:micro} presents a comparison of predictions
of those Skyrme parameterizations which satisfy the macroscopic constraints
with further microscopic and observational constraints. The results are
discussed and summarized in Sec.~\ref{sec:disc} and conclusions are presented in
Sec.~\ref{sec:concl}.

\section{\label{sec:skyrme}Skyrme Models}
Since the original work by Skyrme in the fifties \cite{skyrme1965} and the 
Vautherin and Brink \cite{vautherin1972} parameterization of the original
interaction in early seventies, considerable effort has been invested
in the application of this density dependent effective interaction both to
ground state properties of finite nuclei and to nuclear matter in the framework
of the mean-field Hartree-Fock approximation (see
e.g. \cite{bender2003,stone2007} for recent reviews). The advantage of the
structure of the Skyrme density functional is that it allows analytical
expression of all variables characterising infinite nuclear matter
\cite{chabanat1997,dutra2008,dutra2010}. Such structure can also be constructed
from non-relativistic versions of the relativistic point-coupling models
\cite{greiner2003,lourenco2010,sulaksono2011}. In the following, we introduce
the various physical quantities and give expression for each in terms of the
Skyrme parameters. The general expression for the energy per particle of
infinite ANM, defined in terms of the energy density $\mathcal{E}$ and particle
number density $\rho$, is given 
\begin{eqnarray}
E = \frac{\mathcal{E}}{\rho} &=&
\frac{3\hbar^2}{10M}\left(\frac{3\pi^2}{2}\right)^{2/3}\rho^{2/3}H_{5/3}
+ \frac{t_0}{8}\rho[2(x_0+2)-(2x_0+1)H_2] \nonumber \\
&+& \frac{1}{48}\sum_{i=1}^{3}t_{3i}\rho^{\sigma_{i}+1}
[2(x_{3i}+2)-(2x_{3i}+1)H_2]
+
\frac{3}{40}\left(\frac{3\pi^2}{2}\right)^{2/3}\rho^{5/3}\left(aH_{5/3}+bH_{8/3}
\right)\nonumber \\
&+&\frac{3}{40}\left(\frac{3\pi^2}{2}\right)^{2/3}\rho^{5/3+\delta}\left[
t_4(x_4+2)H_{5/3}- t_4(x_4+\frac{1}{2})H_{8/3}\right]\nonumber \\
&+&\frac{3}{40}\left(\frac{3\pi^2}{2}\right)^{2/3}\rho^{5/3+\gamma}\left[
t_5(x_5+2)H_{5/3}+ t_5(x_5+\frac{1}{2})H_{8/3}\right],
\label{densityenergy}
\end{eqnarray}
with
\begin{eqnarray}
a&=&t_1(x_1+2)+t_2(x_2+2)\mbox{,} \label{eq:a}\\
b&=&\frac{1}{2}\left[t_2(2x_2+1)-t_1(2x_1+1)\right],\quad\mbox{and} 
\label{eq:b} \\
H_n(y)&=&2^{n-1}[y^n+(1-y)^n],
\end{eqnarray}
where $y=Z/A$ is the proton fraction.
Eq.~(\ref{densityenergy}) includes the summation over index $i$ in the third
term introduced by Agrawal {\it et al.} \cite{agrawal2006} and additional terms
involving $t_4$, $x_4$, and $t_5$, $x_5$, used by Chamel {\it et al.}
\cite{chamel2009}. The great majority of the parameterizations referred to in
this work do not include these terms. Parameterization without (with) these
additional terms are regarded as ``standard'' (``non-standard'') in this paper.
We note that there are several other parameter sets which parameterize the
density dependence of the Skyrme functional in non-standard ways
\cite{waroquier1979,liu1991,cochet2004,margueron2009}, different from those
considered here. These forces have been reported to have problems at higher
density nuclear matter \cite{agrawal2006} and have not been included in the
present study.    

All quantities referred to in this work have been obtained based on 
Eq.~(\ref{densityenergy}) and are given below.

Eq.~(\ref{densityenergy}) leads to an in-medium effective nucleon mass $M^*$ in 
ANM
\begin{eqnarray}
M^* &=& M\left\{ H_{5/3}+
\frac{1}{4}\frac{M}{\hbar^2}\rho\left[\left(a+t_4(x_4+2)\rho^\delta +
t_5(x_5+2)\rho^\gamma\right)H_{5/3}\right.\right.\nonumber \\
&+& \left.\left.\left(b - t_4(x_4+\frac{1}{2})\rho^\delta +
t_5(x_5+\frac{1}{2})\rho^\gamma\right)H_{8/3}\right] \right\}^{-1},
\end{eqnarray}
with $M$ being the free nucleon mass.

The pressure, defined as P=$\rho^{2}
\frac{\partial(\mathcal{E}/\rho)}{\partial\rho}$, is given as 
\begin{eqnarray}
P&=& \frac{\hbar^2}{5M}\left(\frac{3\pi^2}{2}\right)^{2/3}\rho^{5/3}H_{5/3}
+\frac{t_0}{8}\rho^2[2(x_0+2)-(2x_0+1)H_2]\nonumber \\
&+&\frac{1}{48}\sum_{i=1}^{3}t_{3i}(\sigma_i+1)\rho^{\sigma_i+2}
[2(x_{3i}+2)-(2x_{3i}+1)H_2]
+\frac{1}{8}\left(\frac{3\pi^2}{2}\right)^{2/3}\rho^{8/3}\left(aH_{5/3}+bH_{8/3}
\right) \nonumber \\ 
&+&\frac{1}{40}\left(\frac{3\pi^2}{2}\right)^{2/3}(5+3\delta)\rho^{\frac{8}{3}
+\delta}\left[t_4(x_4+2)H_{5/3} -t_4(x_4+\frac{1}{2})H_{8/3}\right] \nonumber \\
&+&\frac{1}{40}\left(\frac{3\pi^2}{2}\right)^{2/3}(5+3\gamma)\rho^{\frac{8}{3}
+\gamma}\left[t_5(x_5+2)H_{5/3} +t_5(x_5+\frac{1}{2})H_{8/3}\right].
\label{eq:pressnr}
\end{eqnarray}

The volume incompressibility of ANM at saturation density is calculated as 
derivative of pressure with respect to number density $\rho$
\begin{eqnarray}
K &=& 9\rho^2\left(\frac{\partial^2\mathcal{E}/\rho}{\partial\rho^2}\right)
\nonumber \\
&=& 9\left(\frac{\partial P}{\partial\rho}\right)\nonumber \\
&=&\frac{3\hbar^2}{M}\left(\frac{3\pi^2}{2}\right)^{2/3}\rho^{2/3}H_{5/3}
+\frac{9t_0}{4}\rho[2(x_0+2)-(2x_0+1)H_2]\nonumber \\
&+&\frac{3}{16}\sum_{i=1}^{3}t_{3i}(\sigma_i+1)(\sigma_i+2)\rho^{\sigma_i+1}
[2(x_{3i}+2)-(2x_{3i}+1)H_2]\nonumber \\
&+&3\left(\frac{3\pi^2}{2}\right)^{2/3}\rho^{5/3}(aH_{5/3}+bH_{8/3})\nonumber \\
&+&\frac{3}{40}\left(\frac{3\pi^2}{2}\right)^{2/3}(5+3\delta)(8+3\delta)\rho^{
\frac{5}{3}+\delta}
[t_4(x_4+2)H_{5/3}-t_4(x_4+\frac{1}{2})H_{8/3}] \nonumber \\
&+&\frac{3}{40}\left(\frac{3\pi^2}{2}\right)^{2/3}(5+3\gamma)(8+3\gamma)\rho^{
\frac{5}{3}+\gamma}
[t_5(x_5+2)H_{5/3}+t_5(x_5+\frac{1}{2})H_{8/3}].
\label{eq:incomp}
\end{eqnarray}
Finally, the third derivative of the energy per particle in ANM at saturation 
density, also called the skewness coefficient, is expressed as
\begin{eqnarray}
Q &=& 27\rho^3\left(\frac{\partial^3 \mathcal{E}/\rho}
{\partial \rho^3}\right)\nonumber \\
&=&\frac{12\hbar^2}{5M}\left(\frac{3\pi^2}{2}\right)^{2/3}\rho^{2/3}H_{5/3}
\nonumber \\
&+&\frac{9}{16}\sum_{i=1}^{3}t_{3i}\sigma_i(\sigma_i+1)(\sigma_i-1)
\rho^{\sigma_i+1}[2(x_{3i}+2)-(2x_{3i}+1)H_2] \nonumber \\
&-&\frac{3}{4}\left(\frac{3\pi^2}{2}\right)^{2/3}\rho^{5/3}(aH_{5/3}+bH_{8/3})\nonumber \\
&+&\frac{3}{40}\left(\frac{3\pi^2}{2}\right)^{2/3}
(2+3\delta)(5+3\delta)(3\delta-1) \rho^{\frac{5}{3}+\delta}[t_4(x_4+2)H_{5/3}
-t_4(x_4+\frac{1}{2})H_{8/3}] \nonumber \\
&+&\frac{3}{40}\left(\frac{3\pi^2}{2}\right)^{2/3}
(2+3\gamma)(5+3\gamma)(3\gamma-1) \rho^{\frac{5}{3}+\gamma}[t_5(x_5+2)H_{5/3}
+t_5(x_5+\frac{1}{2})H_{8/3}].\nonumber
\\
\label{eq:skewness}
\end{eqnarray}

In a special case for the SNM ($y=1/2$), the expressions
(\ref{densityenergy} - \ref{eq:skewness}) simplify in that
$H_{\rm n} = 1 $ and we obtain  energy per particle $E_{\rm SNM}(\rho)$, the 
incompressibility $K_{\rm SNM}(\rho)$ and skewness  $Q_{\rm SNM}(\rho)$. When
calculating properties of symmetric matter at the saturation density $E_{\rm
SNM}(\rho_{\rm o})= E_{\rm o}$, the incompressibility $K_{\rm SNM}(\rho_{\rm
o})= K_{\rm o}$ and the skewness $Q_{\rm SNM}(\rho_{\rm o})= Q_{\rm o}$, the
second and third derivatives of the energy density with respect to number
density are taken at $\rho = \rho_{\rm o}$. Obviously, the first derivative, the
pressure, is equal to zero at $\rho_{\rm o}$.
   
One of the key properties of nuclear matter is the symmetry energy, particularly
important in modelling nuclear matter and finite nuclei because it probes the
isospin part of the Skyrme interaction. It is defined as
\begin{eqnarray}
\mathcal{S}(\rho) &=&\frac{1}{8}\frac{\partial^2({\mathcal E}/\rho)}{\partial
y^2}
\big|_{\rho,y={1/2}}\nonumber \\ &=&\frac{\hbar^2}{6M}
\left(\frac{3\pi^2}{2}\right)^{2/3}\rho^{2/3}-\frac{t_0}{8}(2x_0+1)\rho 
-\frac{1}{48}\sum_{i=1}^{3}t_{3i}(2x_{3i}+1)\rho^{\sigma_i+1}\nonumber \\
&+&\frac{1}{24}\left(\frac{3\pi^2}{2}\right)^{2/3}[a+4b]\rho^{5/3}
-\frac{1}{8}\left(\frac{3\pi^2}{2}\right)^{2/3}t_4x_4
\rho^{\frac{5}{3}+\delta}\nonumber \\
&+&\frac{1}{24}\left(\frac{3\pi^2}{2}\right)^{2/3}t_5(5x_5+4)\rho^{\frac{5}{3}
+\gamma}.
\label{eq:esym}
\end{eqnarray}
In SNM, it is customary to define four quantities, $J$=$\mathcal{S}$($\rho_{\rm
o}$), $L$, $K_{\rm sym}$ and $Q_{\rm sym}$ related to the symmetry energy and
its derivatives evaluated at the saturation density $\rho_{\rm o}$. $L$, the
slope of $\mathcal{S}$, is given by
\begin{eqnarray}
L&=&3\rho_{\rm o}\left(\frac{\partial
\mathcal{S}}{\partial\rho}\right)_{\rho=\rho_{\rm o}}\nonumber \\
&=&\frac{\hbar^2}{3M}\left(\frac{3\pi^2}{2}\right)^{2/3}\rho_{\rm o}^{2/3}
-\frac{3t_0}{8}(2x_0+1)\rho_{\rm o} -\frac{1}{16}\sum_{i=1}^{3}t_{3i}
(2x_{3i}+1)(\sigma_i+1)\rho_{\rm o}^{\sigma_i+1}\nonumber \\
&+&\frac{5}{24}\left(\frac{3\pi^2}{2}\right)^{2/3}(a+4b)\rho_{\rm o}^{5/3}
-\frac{1}{8}\left(\frac{3\pi^2}{2}\right)^{2/3}(5+3\delta)t_4x_4\rho_{\rm
o}^{\frac{5}{3}+\delta} \nonumber \\
&+&\frac{1}{24}\left(\frac{3\pi^2}{2}\right)^{2/3}(5+3\gamma)t_5(5x_5+4)\rho_{
\rm o}^{\frac{5}{3}+\gamma}.
\label{eq:slope}
\end{eqnarray}
The curvature of the symmetry energy $\mathcal{S}$ at saturation density in SNM 
is sometimes called $K_{\rm sym}$, the symmetry incompressibility. It should not
be confused with $K_\tau$, which is the isospin incompressibility, defined in
Eqs.~(\ref{isok}) and ~(\ref{ktau}). $K_{\rm sym}$ is given by
\begin{eqnarray}
K_{\rm sym}&=&9\rho_{\rm o}^2\left(\frac{\partial^2
\mathcal{S}}{\partial\rho^2}\right)_{\rho=\rho_{\rm o}}\nonumber \\
&=& -\frac{\hbar^2}{3M}\left(\frac{3\pi^2}{2}\right)^{2/3}\rho_{\rm o}^{2/3}
-\frac{3}{16}\sum_{i=1}^{3}t_{3i}(2x_{3i}+1)(\sigma_i+1)\sigma_i\rho_{\rm
o}^{\sigma_i+1}\nonumber \\
&+&\frac{5}{12}\left(\frac{3\pi^2}{2}\right)^{2/3}(a+4b)\rho_{\rm o}^{5/3}
-\frac{1}{8}\left(\frac{3\pi^2}{2}\right)^{2/3}(5+3\delta)(2+3\delta)t_4x_4\rho_
{\rm o}^{\frac{5}{3}+\delta} \nonumber \\
&+&\frac{1}{24}\left(\frac{3\pi^2}{2}\right)^{2/3}
(5+3\gamma)(2+3\gamma)t_5(5x_5+4)\rho_{\rm o}^{\frac{5}{3}+\gamma}.
\end{eqnarray}
Finally, $Q_{\rm sym}$, the third derivative of the symmetry energy, is
\begin{eqnarray}
Q_{\rm sym}&=&27\rho_{\rm o}^3\left(\frac{\partial^3
\mathcal{S}}{\partial\rho^3}\right)_{\rho=\rho_{\rm o}}\nonumber \\
&=& \frac{4\hbar^2}{3M}\left(\frac{3\pi^2}{2}\right)^{2/3}\rho_{\rm o}^{2/3}
-\frac{9}{16}\sum_{i=1}^{3}t_{3i}(2x_{3i}
+1)(\sigma_i+1)\sigma_i(\sigma_i-1)\rho_{\rm o}^{\sigma_i+1}\nonumber \\
&-&\frac{5}{12}\left(\frac{3\pi^2}{2}\right)^{2/3}(a+4b)\rho_{\rm o}^{5/3}
-\frac{1}{8}\left(\frac{3\pi^2}{2}\right)^{2/3}
(5+3\delta)(2+3\delta)(3\delta-1)t_4x_4\rho_{\rm o}^{\frac{5}{3}+\delta}
\nonumber \\
&+&\frac{1}{24}\left(\frac{3\pi^2}{2}\right)^{2/3}
(5+3\gamma)(2+3\gamma)(3\gamma-1)t_5(5x_5+4) \rho_{\rm o}^{\frac{5}{3}+\gamma}.
\end{eqnarray}

Using the above expressions, the density dependence of the symmetry energy can 
be expanded as a function of $x = (\rho - \rho_{\rm o})/3\rho_{\rm o}$,
\begin{equation}
\mathcal{S}=J+Lx+\frac{1}{2}K_{sym}x^2+ \frac{1}{6}Q_{sym}x^3+\mathcal{O}(x^4).
\label{eq:expansion}
\end{equation}

Similarly, in SNM, the density dependence of the energy per particle $E$, 
Eq.~(\ref{densityenergy}), is sometimes expressed as an expansion in  a Taylor
series around $\rho_{\rm o}$
\begin{equation}
E_{\rm SNM}(\rho) = E_{\rm o} +\frac{1}{2}K_{\rm o}
x^{\rm 2} + \frac{1}{6}Q_{\rm o} x^{\rm 3}+\mathcal{O}(x^{\rm 4}).
\end{equation}
where E$_{\rm o}$ = E$_{\rm SNM}$($\rho_{\rm o}$) is the energy per particle at the saturation density $\rho_{\rm o}$ and
\begin{equation}
K_{\rm o}=9\rho_{\rm o}^{\rm 2}\left(\frac{\partial^{\rm 2}E_{\rm SNM}(\rho)}
{\partial\rho^{\rm 2}}\right)_{\rho=\rho_{\rm o}}
\label{eq:k0}
\end{equation}
and
\begin{equation}
Q_{\rm o}=27\rho_{\rm o}^{\rm 3}\left(\frac{\partial^{\rm 3}E_{\rm SNM}(\rho)}
{\partial\rho^{\rm 3}}\right)_{\rho=\rho_{\rm o}}
\label{eq:q0}
\end{equation}

In  ANM with asymmetry $\beta=(N-Z)/A=(1-2y)$, the energy per 
particle $E$ can be expanded around a new, isospin dependent, saturation
density
$\rho_{\rm o}(\beta) \sim \rho_{\rm o} (1 - 3(L/K_{\rm o})\beta^{\rm 2})$
\cite{vidana2009}:
\begin{equation}
E_{\rm ANM}(\rho,\beta)=E_{\rm o}(\rho_{\rm o}(\beta))+\frac{K_{\rm
o}(\rho_{\rm o}(\beta))}{2}\left(\frac{\rho-\rho_{\rm o}(\beta)}{3\rho_{\rm
o}(\beta)}\right)^2+\frac{Q_{\rm o}(\rho_{\rm o}(\beta))}{6}\left(\frac{\rho-
\rho_{\rm o}(\beta)}{3\rho_{\rm o}(\beta)}\right)^3 + \mathcal{O}(\beta^{\rm 4})
\label{erhobeta}
\end{equation}
where the expansion coefficients are given as
\begin{eqnarray}
E_{\rm o}(\rho_{\rm o}(\beta)) &=& E_{\rm o} + J\beta^{\rm 2} +
\mathcal{O}(\beta^{\rm 4}), \label{ebeta}  \\
K_{\rm o}(\rho_{\rm o}(\beta)) &=& K_{\rm o} + \left(K_{\rm sym} - 6L
-\frac{Q_{\rm o}}{K_{\rm o}}L\right)\beta^{\rm 2} + \mathcal{O}(\beta^{\rm 4}),
\qquad \mbox{and} \label{kbeta}\\
Q_{\rm o}(\rho_{\rm o}(\beta)) &=& Q_{\rm o} + \left(Q_{\rm sym} -
9L\frac{Q_{\rm o}}{K_{\rm o}}\right)\beta^{\rm 2} + \mathcal{O}(\beta^{\rm 4}).
\label{qbeta}
\end{eqnarray}
The coefficient of the second term in Eq.~(\ref{kbeta}) 
\begin{equation}
K_{\tau,v}=\left(K_{\rm sym} - 6L - \frac{Q_{\rm o}}{K_{\rm o}}L\right)
\label{isok}
\end{equation}
determines the isospin dependence of incompressibility at saturation density $\rho_{\rm o}(\beta)$. 
Strictly it is the volume part $K_{\tau, v}$ of the isospin incompressibility
$K_\tau$ Eq.~(\ref{ktau}), which plays an important role in analysis of data
from giant monopole resonance. It does not include surface
effects, as discussed in Secs.~\ref{sub:snm} and \ref{sub:mix}.

\section{\label{sec:macro}Macroscopic constraints}
It is important to keep in mind that different Skyrme parameterizations were 
often constructed with emphasis on a certain selection of data on finite nuclei.
For example, the BSk family were fitted to experimental nuclear masses, SkM* 
to binding energies of finite nuclei and actinide fission barriers, the SkI
family to isotope shifts in the Pb region, and the SLy family to properties of
neutron matter, neutron stars and the ground state variables of neutron-heavy
nuclei. Although all Skyrme forces are usually fitted to reproduce well the
saturation energy and density of symmetric nuclear matter, they differ
significantly in other characteristics of symmetric and pure neutron matter, in
particular their density dependence.

We examine in this section eleven constraints on properties of nuclear matter,
out of which four are related to SNM, two to PNM,
and five involving both SNM and PNM. The constraints are
listed in Table~\ref{tab:listcon}.

\subsection{\label{sub:snm}Symmetric nuclear matter}
Infinite nuclear matter, composed of the same number of protons and neutrons 
without Coulomb interaction, does not
exist in nature. Nevertheless, it has become an important theoretical laboratory
for the investigation of physical quantities relevant for the modelling of heavy
nuclei and nuclear matter in astrophysical compact objects. As stated above, the
saturation density $\rho_{\rm o}$ and the binding energy per nucleon $E_0 =
\mathcal{E}/\rho_{\rm o}$ are reasonably well established. In this work we focus
on two other physical quantities of SNM at saturation density: The
incompressibility $K_{\rm o}$ and the skewness coefficient $Q_{\rm o}$.
We note that some authors use $K'=-Q_{\rm o}$ as the skewness coefficient 
(see Eq.~(\ref{eq:skewness})).

The determination of $K_{\rm o}$, and other related parameters of nuclear matter 
from experimental data on Giant Monopole Resonance (GMR) on finite nuclei, has
been a long-standing problem (see e.g \cite{blaizot1980,shlomo1993,blaizot1995})
which has not been fully resolved to this day \cite{pearson2010}.

There are basically two procedures that can be used to extract the information. 
One is to analyse the $A$ dependence of the compression modulus of a finite
nucleus $K_{\rm A}$
\begin{equation}
K_{\rm A}=(M/\hbar^{\rm 2})<R^2>E^{\rm 2}_{\rm GMR}
\end{equation}
obtained, using sum-rule arguments, from the measured energy of Giant Monopole 
Resonance $E_{\rm GMR}$ in spherical nuclei \cite{blaizot1980}. $M$ is the
nucleon mass and $R$ is the rms \textit {matter} radius of the nucleus with mass
number $A$. Using the leptodermous expansion of the energy per particle in the
droplet model (the mass formula), an expression for $K_{\rm A}$ can be found
\cite{blaizot1980},
\begin{eqnarray}
K_{\rm A}& = & K_{\rm vol} + K_{\rm surf}A^{-1/3} + K_{\rm curv}A^{-2/3}
\nonumber \\
          & + & K_{\tau}\beta^{\rm 2} + K_{\rm coul} \frac{Z^{\rm 2}}{A^{\rm
4/3}} + \cdots ,
\label{ka}
\end{eqnarray}
where the isospin incompressibility $K_\tau$ consists of two components 
\cite{blaizot1980,treiner1981,nayak1990}
\begin{equation}
K_\tau=K_{\tau,v}+K_{\tau,s}A^{-1/3}
\label{ktau}
\end{equation}
where $K_{\rm \tau,v}$ ($K_{\tau,s}$ ) determine the volume (surface) isospin 
incompressibility. $K_{\rm vol}$, $K_{\rm surf}$, $K_{\rm curv}$, $K_{\tau}$ and
$K_{\rm coul}$ are second derivatives of the coefficients of the volume,
surface, curvature, isospin, and Coulomb terms with respect to the radial
coordinate of the nucleus in the mass formula. If the $K_{\rm vol}$ term is
identified as the incompressibility of infinite nuclear matter $K_{\rm o}$, then
a link between $K_{\rm A}$ and $K_{\rm o}$ can be used to determine $K_{\rm o}$.
However, it has been argued that the macroscopic analysis suffers from
potentially serious drawbacks, including the uncertainty of the validity of the
$K_{\rm vol} \approx K_{\rm o}$ assumption, the weak $A$ dependence, limitations
of Eq.~(\ref{ka}) to small oscillations, questions about convergence of the
expansion Eq.~(\ref{ka}), and consideration of possible anharmonicities of the
breathing mode, especially for light nuclei \cite{blaizot1995}.

Another route is to rely on microscopic calculations within a Hartree-Fock 
mean-field approximation for static properties and RPA for excitations, and to
use the same model framework for calculation of the SNM. Such an approach allows
a consistent determination of the breathing mode energy  $E_{\rm GMR}$ and
parameters of nuclear matter within the same framework. However, the microscopic
approach has the disadvantage that the results are dependent on the choice of
the effective interaction used in the models. The macroscopic approach is based 
only on the assumption that the liquid drop model of the nucleus is valid and
that the leptodermous expansion (\ref{ka}) converges reasonably fast. 

The value of $K_{\rm o}$ most frequently used today is based on microscopic 
analysis of GMR data. Youngblood {\it et al.} \cite{youngblood1999} used 
measured E0 strength distribution in  $^{\rm 40}$Ca, $^{\rm 90}$Zr, $^{\rm
116}$Sn,  $^{\rm 144}$Sm and  $^{\rm 208}$Pb and Gogny interaction based
on calculations by Blaizot {\it et al.} \cite{blaizot1995}, which took into 
account pairing and anharmonicity in lighter nuclei. The deduced value of
$K_{\rm o}$= 231$\pm$5 MeV. This result was in good agreement with the value
obtained by Farine {\it et al.} \cite{farine1997}  $K_{\rm o}$ = 240 MeV with
generalized Skyrme forces. Myers
and \'{S}wi\c{a}tecki \cite{myers1998} used a model based on a semiclassical 
Thomas-Fermi approximation with a short range Yukawa effective force between
fermions. They obtained the value $K_{\rm o}$= 234 MeV  by fitting nuclear
binding energies and diffuseness to experimental data. Later, Col\`o and
co-workers \cite{colo2004} argued that there is not a unique relationship
between the value of  $K_{\rm o}$ associated with an effective force and the GMR
energy predicted by that force in non-relativistic models. They built a new
class of Skyrme forces and found that 230$<K_{\rm o}<$250 MeV. Agrawal {\it et
al.} addressed the well-known discrepancy between predictions of the value of
$K_{\rm o}$ in non-relativistic and relativistic models (see \cite{agrawal2003}
and references therein) and concluded that the discrepancy, thought to be $\sim$
20\%, can be much smaller for an appropriate choice of effective interactions.
Todd-Rutel and Piekarewicz \cite{todd2005}, using a new relativistic model
FSUGold with two new parameters, causing softening both of the EoS of
SNM and the density dependence of the symmetry energy, predicted $K_{\rm o}$=
230 MeV in RMF calculations. However, new data on GMR in Sn isotopes
\cite{li2007,li2010} re-opened the question of  $K_{\rm o}$, as models which
successfully predicted GMR energies in $^{\rm 90}$Zr, $^{\rm 144}$Sm and  $^{\rm
208}$Pb could not reproduce GMR energies reported for $^{\rm 112-124}$Sn. We
note that the new GMR energies for $^{\rm 112,116,124}$Sn do not agree (just
outside errors) with previous data \cite{youngblood2004, lui2004}. It has been
suggested that the new measurement by Li {\it et al.} offered cleaner spectra
with less need for subtraction of background, which may possibly help to
understand this discrepancy \cite{li2010}. Col\`o {\it et al.} \cite{colo2008}
tried to interpret the new data employing a self-consistent quasiparticle
random-phase approximation model with Hartree-Fock-Bogoliubov basis, the Skyrme
interaction and density dependent pairing.  They reproduced the new
GMR energies in $^{\rm 112-120}$Sn isotopes using SkM* Skyrme force and surface
pairing force and found that the effect of pairing on $K_{\rm o}$ is very small.
However, the value of $K_{\rm 0}$ extracted from the fit to Sn isotopes is about
10\% smaller than the one obtained from the fit to  $^{\rm 208}$Pb, 230-240 MeV.
Similar $\sim$10\% discrepancy between $K_{\rm 0}$ from fits to Sn and Pb GMR
energies has been reported in RMF calculations \cite{vretenar2003}. So, the
puzzle of GMR in Sn nuclei remains open \cite{piekarewicz2010}. Keeping in mind
the current unresolved situation of GMR experiments and theory, and that all
the extracted values are likely to be rather model dependent, we choose the
following constraint for $K_{\rm o}$ that we refer to as the SM1 constraint, 
\begin{eqnarray}
{\bf SM1} \,:\, \,\,\,\,  K_{\rm o} = 230 \pm 30\, {\rm  MeV}
\end{eqnarray}

To our knowledge, there has been only one attempt to constrain the third 
derivative of the energy per particle with respect to density, the skewness
$K^\prime = -Q_{\rm o}$ (see Eq.~(\ref{eq:skewness})) which is the next order
fluctuation around the saturation density in the expansion \cite{farine1997} 
\begin{equation}
\mathcal{E}(\rho)/{\rho} \approx E_{\rm o} + (K_{\rm o}/2) x^2 -(K'/6) x^3
+ \cdots. 
\end{equation}

Constraints on the skewness coefficient, which differs from author to author 
by a minus sign \cite{farine1997,piekarewicz2009}, are relatively imprecise.
Farine {\it et al.} \cite{farine1997} tried to find acceptable values of
$K^\prime$ by self-consistent analysis of breathing mode data using a selection
of Skyrme forces. They found a subtle correlation between $K_{\rm o}$ and
$K^\prime$. We adopt here their value of $K^\prime$ as constraint SM2,
\begin{eqnarray}
{\bf SM2}\,:\,\,\,\,\, K'= 700 \pm 500\,{\rm  MeV}.
\end{eqnarray}
The wide range of $K^\prime$ compensates for the fact that Farine {\it et al.} 
\cite{farine1997} used  $K_{\rm o}$=215 $\pm 15$ MeV, 7\% lower than our
choice. 

Limits on the pressure-density relationship in SNM and PNM and its curvature 
can be obtained from analysis of experimental data on the motion of ejected
matter in energetic nucleus-nucleus collisions. Recently, measurements of the
particle flow in collisions of $^{197}$Au nuclei at incident kinetic
energy per nucleon varying from about $0.15$ to $10$ GeV were analysed
\cite{danielewicz2002}. The authors extrapolated available data
\cite{brill1996} for pressure at about $2 \rho_{\rm o}$ to higher densities, as
well as to zero temperature. The results give limits on pressure as a function
of density (see Fig.~\ref{fig:sm3}) which comprise the constraint SM3,
\begin{eqnarray}
{\bf SM3}\,: \,\,\,\,\, P\,\,\, {\rm vs}\,\,\, \frac{\rho}{\rho_{\rm o}}(y=0.5)
\Longrightarrow Flow\,\, Experiment.
\end{eqnarray}

Recent experimental investigation of kaon production in HIC 
produces a further constraint on pressure at a lower density region
$1.2\le\rho\le2.2$ fm$^{-3}$ \cite{lynch2009,fuchs2006} (see
Fig.~\ref{fig:sm4}), here named SM4,
\begin{eqnarray}
{\bf SM4}\,: \,\,\,\,\, P\,\, {\rm vs}\,\, \frac{\rho}{\rho_{\rm o}}(y=0.5) 
\Longrightarrow Kaons + GMR\,\,Experiments.
\end{eqnarray}

\subsection{\label{sub:pnm} Pure Neutron Matter} 
The EoS of PNM is of a particular interest, because PNM is a realistic first 
approximation to the baryonic matter that composes neutron stars. Most 
properties of neutron stars cannot be studied in terrestrial laboratories and
theoretical models, based on effective forces, must be used. However, at low
densities, experiments with cold Fermi atoms yield information on strongly
interaction fluids, very similar to the low-density neutron matter at neutron
star crusts \cite{gezerlis2008}. Different density regimes can be tuned by the
magnitude of the neutron Fermi momentum $k_{\rm F}$ relative to the effective
range $r_{\rm o}$ of the NN interaction in the system \cite{schwenk2005}. The
ground state energy per particle, the EoS, can be expressed as
\begin{equation} 
\frac{E_{\rm PNM}}{E_{\rm PNM}^{o}} = \xi, 
\label{unitarity} 
\end{equation} 
where $E_{\rm PNM}$ is the energy per particle in Eq.~(\ref{densityenergy})
with $H_{\rm n}=2^{\rm n-1}$.  $E_{\rm PNM} = {\mathcal E_{\rm PNM}}/{\rho}$
and $E_{\rm PNM}^{o}=3\hbar^2k_{\rm F}^{2}/10M$, with $M$ being the mass of the
nucleon. In the dilute degenerate Fermi gas regime, $k_F r_{\rm o} \ll $ 1,
$\xi$ is a constant \cite{carlson2003a}. This restricts the density below about 
10$^{\rm -3}\rho_{\rm o}$, the density at which neutrons become unbound in
neutron stars. At higher densities, below $\sim$ 0.1$\rho_{\rm o}$, where  $k_F
r_{\rm o} \approx $ 1, $\xi$ has to be replaced by a system dependent function
$\xi$(k$_{\rm F}$, r$_{\rm o}$). In this work we adopt the expression
$E_{PNM}/E_{PNM}^{o}$ by Epelbaum {\it et al.} \cite{epelbaum2009}, based on
next-to-leading order in lattice chiral effective field theory (NLO$_{\rm 3}$),
and including corrections due to finite scattering length, nonzero effective
range, and higher order corrections,
\begin{eqnarray}
\frac{E_{PNM}}{E_{PNM}^{o}} = \xi-\frac{\xi_1}{k_Fa_{\rm o}} + c_1k_Fr_{\rm o} +
c_2k_F^2m_\pi^{-2} + c_3k_F^3m_\pi^{-3}+\cdots, 
\label{eq:106} 
\end{eqnarray}
where $m_{\rm \pi}$ is the pion mass. The dimensionless universal constant $\xi$
has been determined from trapped cold atom experiments with $^{\rm 6}$Li and
$^{\rm 40}$K, which yield a variety of values: $0.32_{-13}^{+10}$
\cite{bartenstein2004}, $0.51(4)$ \cite{kinast2005}, $0.46_{-05}^{+12}$
\cite{stewart2006}, and $0.39(2)$ \cite{luo2008}.  Values of $\xi_1$ in the
literature are in the range 0.8 - 1.0 (\cite{epelbaum2009} and references
therein). Epelbaum {\it et al.}, using a simple Hamiltonian and only few
particles in their system, took  $\xi=0.31$ and $\xi_{1}=0.81$ and fitted two
sets of constants c$_{\rm 1}$, c$_{\rm 2}$ and c$_{\rm 3}$: set 1 (0.27, -0.44,
0.0) and set 2 (0.17, 0.0, -0.26), and obtained a very similar quality fits to
their NLO$_{\rm 3}$. We construct a constraint on energy per particle of PNM in
the range of densities 0.01 - 0.1$\rho_{\rm o}$ shown in Fig.~\ref{fig:pnm1}
\begin{eqnarray}
{\bf PNM1}\,:\,\,\,\,\, \frac{\mathcal E_{PNM}}{\rho} ({\rm MeV})\,\,\,{\rm
vs}\,\,\, \rho.
\end{eqnarray}
Two shaded areas are based on Eq.~(\ref{eq:106}) with $\xi_{1}=0.81$ and set 1 
(red dashed line) or set 2 (green dashed-dotted line). The area inside the blue
solid line is based on Eq.~(\ref{unitarity}), the unitary limit. The boundaries
of all three areas are calculated by taking 0.2$ < \xi < $0.6, which allows for
the spread in experimental values.       
 
It is clear that our PNM1 constraint is consistent with a relatively large range
of extrapolated experimental $\xi$ values. Very recently, after our work was
completed, more accurate calculations and measurements were reported. The new
limits on $\xi$ are 0.37-0.38 \cite{ku2011} and 0.38-0.41 \cite{gandolfi2011a}.
$\xi_{\rm 1}$, related to the contact parameter in unitary Fermi gases, has been
calculated  to be $\xi_{\rm 1}$=0.9 (\cite{gandolfi2011a} and references
therein). These new data will be considered in future development of the PNM1
constraint.

New theoretical calculations also provide constrains on the EoS of low-density 
PNM. Gezerlis and Carlson (\cite{gezerlis2010} and references therein) compare
their EoS, obtained using the Quantum Monte Carlo techniques, with results of
other model calculations in their Figs. 3 and 4. Although we still keep the
PNM1 as selector of the low-density neutron matter 
EoS in this paper, we have constructed a band, representing the boundaries of 
theoretical predictions, in the same $\rho/\rho_{\rm o}$ range used in the PNM1
constraint. This band is considerably narrower than the band extracted from the
cold atoms data. It is shown in Fig.~\ref{fig:pnm1star}a, and merged with the
PNM1 constraint in Fig.~\ref{fig:pnm1star}b. 

In the high density region, analysis of HIC data \cite{danielewicz2002} provides
a constraint on the pressure-density relation in PNM.
\begin{eqnarray}
{\bf PNM2}: P\,\,\,{\rm vs}\,\,\,\frac{\rho}{\rho_{\rm o}}(y=0) 
\Longrightarrow Flow\,\, Experiment.
\end{eqnarray}
The constraint is illustrated in Fig.~\ref{fig:pnm2}.
\subsection {\label{sub:mix}Constraints involving both SNM and PNM}
Nuclear symmetry energy, in particular its density dependence, has received 
considerable attention during the last decade. It produces information about the
isospin dependence of nuclear forces which is equally important in nuclear
matter and finite nuclei. The current empirical values
of $J$, the symmetry energy at saturation density, as predicted by different 
models, vary around the value extracted from the up-to-date finite range liquid
droplet model $J=32.5$ MeV \cite{moller2012}. The data for $\mathcal{S}(\rho)$
come from several sources: HIC \cite{lynch2009,tsang2009, klim2007,kohley2011}, 
Pygmy Dipole Resonances (PDR) \cite{klim2007,wieland2009,carbone2010}, Isobaric
Analog States (IAS) \cite{dan2011} and numerous microscopic calculations. A
systematic difference exists between predictions of HF (Skyrme) and RMF models
\cite{bali2008,stone2007} spanning the range $27<J<38$ MeV.  This large
uncertainty stems, in part, from limited experimental knowledge and
understanding of the isospin dependence of the nucleon-nucleon interaction and,
in particular, the PNM EoS. Taking into account some more recent data
\cite{tsang2012} we define this constraint as
on $J$,
\begin{eqnarray}
{\bf MIX1} \,: \,\,\,\,\,30<J<35\, {\rm  MeV}.
\end{eqnarray}  

The density dependence of the symmetry energy, especially at
super-nuclear densities, has direct relevance for modelling neutron stars
\cite{bali2008,stone2003} and is closely related to studies of neutron matter
radii and the neutron skin in neutron-heavy nuclei \cite{brown2000,typel2001}.
In contrast to the expansion of the energy per particle, in which the term 
containing the first derivative vanishes, the expansion of $\mathcal{S}(\rho)$
Eq.~(\ref{eq:expansion}) contains a first order correction $L$ at
$\rho=\rho_{\rm o}$. $L$ becomes an important bulk quantity that determines most
of the behavior of $\mathcal{S}(\rho)$ in the vicinity of $\rho_{\rm o}$. The
empirical determination of $L$ is, as for several other bulk quantities,
indirect. The very recent constraint, based on the empirical  MSL model with the
MSL0 Skyrme-like interaction, is based on analysis of isospin
diffusion and double neutron/proton ratio in heavy-ion collisions at
intermediate energies, and requires \cite{chen2010}
\begin{eqnarray} 
{\bf MIX2}\,:\,\,\,\,\,L = 58 \pm 18\, {\rm  MeV}.
\end{eqnarray}
We comment that this value is lower than the previously accepted $L = 88 \pm 25$
MeV \cite{chen2005} derived by the same group but considering only the isospin
diffusion data and standard Skyrme forces. This lower value of $L$ is supported
by findings of Newton and Li, \cite{newton2009a}, who used the correlation
between the gravitational binding energy of a low mass neutron star PSR
J0737-3039B and the slope of the nuclear symmetry energy at 1--2 times the
nuclear saturation density. This correlation leads to an upper limit $L \leq 70$
MeV. It is also consistent with the value $L = 49.9$ MeV determined from the
droplet model \cite{myers1996} and closer to the most recent value from Finite 
Range Droplet Model (FRDM) $L = 70 \pm 15$ MeV \cite{moller2012}. Vida\~na {\it
et al.} \cite{vidana2009} found the value of $L$ = 66.5 MeV in their
Bruckner-Hartree-Fock calculation with the Argonne V18 potential supplemented by
a  three-body force of Urbana type. As $J$ and $L$ are correlated, some
investigations produce range of acceptable values for both of these observables
(see e.g. \cite{carbone2010,tsang2012} for a recent summary). 

The next constraint involves the isospin incompressibility $K_{\tau}$ in 
Eq.~(\ref{ka}) for the compression modulus of a finite nucleus with mass $A$.
This  constraint on $K_{\tau}$, which is only dependent on the validity of the 
expansion (\ref{kali}), from experiment was provided by Li {\it et al.}
\cite{li2007, li2010,garg2011} who measured GMR strength distributions for
$^{112-124}$Sn and $^{106,110-116}$Cd isotopes in inelastic scattering of alpha
particles. They used a simplified expression for the compression modulus $K_A$
(as compared to Eq.~(\ref{ka})),
\begin{equation}
K_{\rm A}= K_{\rm vol} + K_{\rm surf}A^{-1/3}
          + K_{\tau}\beta^{\rm 2}
          + K_{\rm Coul} \frac{Z^{\rm 2}}{A^{\rm 4/3}},
\label{kali}
\end{equation}
omitting the higher order terms in $A$ and $\beta$. Li {\it et al.} further
assumed that $K_{\rm surf} = - K_{\rm vol}$ and $K_{\rm vol}$ = $K_{\rm o}$ (the
scaling approximation).  $K_{\rm Coul}$ was taken to be (-5.2 $\pm$ 0.7) MeV.
This value has been derived by  Sagawa {\it et al.} \cite{sagawa2007}, who
investigated a microscopic structure of $K_{\rm Coul}$ and its correlation with
$K_{\rm o}$, using a set of thirteen Skyrme parameterizations and seven RMF
Lagrangians. They found this correlation rather weak and arrived to an average
value of  $K_{\rm Coul}$,  used by Li {\it et al.} This procedure yielded the
value of $K_{\tau}$ = (-550 $\pm$ 100) MeV. 

It is important to realize that the $K_{\tau}$, extracted by Li {\it et al.},
includes both the volume and the surface components of the isospin
compressibility $K_\tau$, Eq.~(\ref{ktau}). It follows that, in order to
compare microscopic model calculation with the constraint, the contribution of
the surface-symmetry term, which is difficult to calculate exactly, must be at
least estimated as well as possible and subtracted from $K_{\tau}$.  Stone {\it
et al.} \cite{stone2012} re-analysed a combined Sn+Cd data by Li {\it et al.} and
Garg under the same conditions and found $K_{\tau}$ = (-595 $\pm$ 154) MeV.
Estimation of $K_{\rm \tau,v}$ and $K_{\rm \tau,s}$ from the currently available
GMR data is not easy. The data show systematic differences, mainly dependent on
methods, used for analysis by different groups. It is therefore necessary to
include some additional constraints on the fit to obtain limits on $K_{\rm
\tau,v}$ and $K_{\rm \tau,s}$. Stone {\it et al.} used two assumptions. First,
they required  that Eq.~(\ref{ktau}) holds and looked for all combinations of
$K_{\rm \tau,v}$ and $K_{\rm \tau,s}$ which would satisfy it. $K_{\rm \tau,v}$
is expected to be negative in line with microscopic calculations. It was varied
in the region of -1200 $ < K_{\rm \tau,v} <$ 0 MeV with $K_{\rm \tau,s}$ in the
range of -1600  $< K_{\rm \tau,s} <$ 1600 MeV. The second assumption was that
the expansion (\ref{ktau}) in terms of $A^{-1/3}$ and $\beta^2$ converges at a
reasonable rate, i.e., no higher order term are significant. They took
\begin{equation}
\frac{K_{\tau,s}A^{-1/3}}{K_{\tau,v}} \le 0.5
\label{limit}
\end{equation}
Simultaneous application of Eq.~(\ref{ktau}) and Eq.~(\ref{limit}), together
with the assumption that $K_{\tau,v}$ is negative, allows limits on $K_{\tau,v}$
to be extracted and taken as a constraint,
\begin{eqnarray}
{\bf MIX3}\,:\,\,\,\,\,-760 \le K_{\tau,v} \le -372\, {\rm  MeV}.
\end{eqnarray}

Corresponding limits on the surface contribution to isospin incompressibility 
are -1110 $\le K_{\tau,s} \le$ 960 MeV. We comment that the condition on the
ratio of the contribution of the surface and volume isospin incompressibility is
rather conservative. Treiner {\it et al.} \cite{treiner1981} estimated the
surface contribution to the isospin incompressibility to be $\sim$20\% of the
volume term in $^{\rm 208}$Pb in the scaling approximation, implying a rapid
convergence of expansion, Eq.~(\ref{ka}). 
We note that other suggested limits on $K_{\rm \tau,v}$  = -370 $\pm$ 120 MeV 
(in notation of the original paper \cite{chen2009} $ K_{\tau,2}^{sat}$) exist in
the literature, but they are calculated, not directly extracted from
experimental data.  Patra et al. \cite{patra2002} estimated  $K_{\rm \tau,v}$
and $K_{\tau,s}$ using a semiclassical relativistic mean-field method with
interaction NL1, NL3 and NLSH. They obtained $K_{\rm \tau,v}$ = -676, -690 and
-794 MeV and  $K_{\rm \tau,v}$ = 1951, 1754 and 1716 MeV for the three
interactions, NL1, NL3 and NLSH, respectively.  

As the next constraint, we used the known effect that at the nuclear surface, 
variation of the difference between proton and neutron densities (neutron skin)
is expected. Danielewicz \cite{danielewicz2003} considered this question and
proposed limits to the reduction of the symmetry energy at $\rho_{\rm o}/2$ in
terms of $\mathcal{S}(\rho_{\rm o}/2)/J$. These limits,
based on a semi-classical Thomas-Fermi model and a comparison of the neutron
skin of $^{\rm 208}$Pb as calculated in this model with values from full
mean-field models, lead to 
\begin{eqnarray} 
{\bf MIX4}\,:\,\,\,\,\, 0.57<\frac{\mathcal{S}(\rho_{\rm o}/2)}{J}<0.83.
\end{eqnarray}

Finally, Piekarewicz \cite{piekarewicz2002,piekarewicz2007} used  a parabolic 
approximation to the EoS and derived an expression for the pressure in pure
neutron matter, related to the slope of the symmetry energy $L$ at saturation
density. Thus the ``symmetry pressure'' $L$, a quantity that influences the
neutron-skin thickness in heavy nuclei, is directly proportional to the pressure
of pure neutron matter as $3P_{PNM}(\rho_{\rm o})/(L\rho_{\rm o}) \approx 1$
\cite{piekarewicz2007}. Considering the uncertainty in the number of terms
included in the expansion, we introduce a range to this constraint as  
\begin{eqnarray}
{\bf MIX5}\,: \frac{3P_{PNM}(\rho_{\rm o})}{L\rho_{\rm o}} = 1 \pm 0.1.
\label{mix5}
\end{eqnarray}
 
In terms of Skyrme model parameters, this constraint can be expressed 
analytically as 
\begin{eqnarray}
\frac{3P_{PNM}(\rho_{\rm o})}{L\rho_{\rm o}} &=& 1
+\frac{1}{L}\sum_{i=1}^{6}T_i,
\label{pnm1+}
\end{eqnarray}
where,
\begin{eqnarray}
T_1 &=& \left(\frac{9\sqrt[3]{4}}{5}-1\right)\frac{\hbar^2}{3M}
\left(\frac{3\pi^2}{2}\right)^{2/3}\rho_{\rm o}^{2/3}
\nonumber \\
T_2 &=& \frac{9t_0}{8}\rho_{\rm o}\nonumber \\
T_3 &=& \frac{3}{16}\sum_{i=1}^{3}t_{3i}(\sigma_i+1)\rho^{\sigma_i+1}\nonumber
\\
T_4 &=& \frac{5}{24}\left(\frac{3\pi^2}{2}\right)^{2/3}
\left[a\left(\frac{9\sqrt[3]{4}}{5}-1\right) 
+ 2b\left(\frac{9\sqrt[3]{4}}{5}-2\right)\right]\rho_{\rm o}^{5/3}\nonumber \\
T_5 &=& \frac{1}{8}\left(\frac{3\pi^2}{2}\right)^{2/3}\left(5+3\delta\right)
\left[\frac{3\sqrt[3]{4}}{5}(1-x_4)+x_4\right]t_4\rho_{\rm
o}^{\frac{5}{3}+\delta}\nonumber \\
T_6 &=& \frac{1}{8}\left(\frac{3\pi^2}{2}\right)^{2/3}\left(5+3\gamma\right)
\left[\frac{9\sqrt[3]{4}}{5}(1+x_5)-\frac{1}{3}(5x_5+4)\right]t_5\rho_{\rm
o}^{\frac{5}{3}+\gamma}.
\label{pnmterms}
\end{eqnarray}
 
\subsection{\label{sub:resmac}Results of application of the macroscopic
constraints}
Predictions of bulk properties of nuclear matter by all Skyrme parameter sets 
are summarized in Table~\ref{tab:237}. Table~\ref{tab:const} details whether a
particular parameter set is consistent (+) or not (-) with each of the eleven
constraints. 

Numerical evaluation of the compliance for individual parameterizations with 
the constraints is given in Table~\ref{tab:const-num}. The criterion for
consistency with a constraint is different for numerical (SM1, SM2 and MIX1-5)
and graphical (SM3, SM4, PNM1 and PNM2) constraints. For the numerical ones, we
define a deviation $Dev$
\begin{equation}
Dev = \frac{Q_{\rm mod}-Q_{\rm const}}{\Delta},
\end{equation}
where $Q_{\mbox{\scriptsize{mod}}}$ is the value of a specific quantity 
calculated in the model, and $Q_{\mbox{\scriptsize{const}}}$ is the central
value of the related constraint. The error associated with $Q_{\rm const}$ is
$\Delta$. If  $|Dev|\le 1$, the model is consistent. In particular, for the
MIX1, MIX3 and MIX4 constraints that are defined by a range in the form $x_1\le
X \le x_2$, we define the central value as $Q_{\rm const}=(x_2+x_1)/2$, and the
error as $\Delta=x_2-Q_{\rm const}=Q_{\rm const}-x_1$.

For the graphic constraints SM3, SM4, PNM1 and PNM2, we consider as consistent a
model that is inside of the corresponding band in $95\%$ or more of the density
region. In this case, the Skyrme interaction is rendered valid.

Table~\ref{tab:nap} shows the number of parameter sets (out of 240) which 
satisfied each constraint. This table shows that no individual constraint is
particularly discriminative. It is more interesting to consider further the
parameter sets which satisfy all constraints. They are surprisingly few in
number -- only {\bf{six}} sets {\bf LNS, NRAPR, Ska25s20, SQMC650, SQMC700, and
SV-sym32} are selected. 

Further, considering the need to choose ranges for some constraints, we looked 
for sets that fell outside the chosen range for only one of the eleven
constraints and by less then 5\% of the closest limit. This procedure yielded
ten more sets {\bf GSkI, GSkII, KDE0v1, MSL0, Ska35s20, SKRA, SkT1, SkT2,
SkT3, and Skxs20}, making a total of {\bf sixteen} Consistent Skyrme
Parameterizations (hereafter CSkP) which satisfy all constraints on properties
of nuclear matter. (We note that sets SkT1a, SkT2a and SkT3a (see
Table~\ref{tab:237}) have the same parameters relating to nuclear matter as
SkT1, SkT2, SkT3 \cite{tondeur1984}, and differ only by a choice of the
spin-orbit functional, the Coulomb exchange term and the fitted pairing
strength. SkT1a, SkT2a and SkT3a are therefore not included separately in the
analysis).

\begin{figure}[ht]
\begin{center}
\includegraphics[scale=0.6]{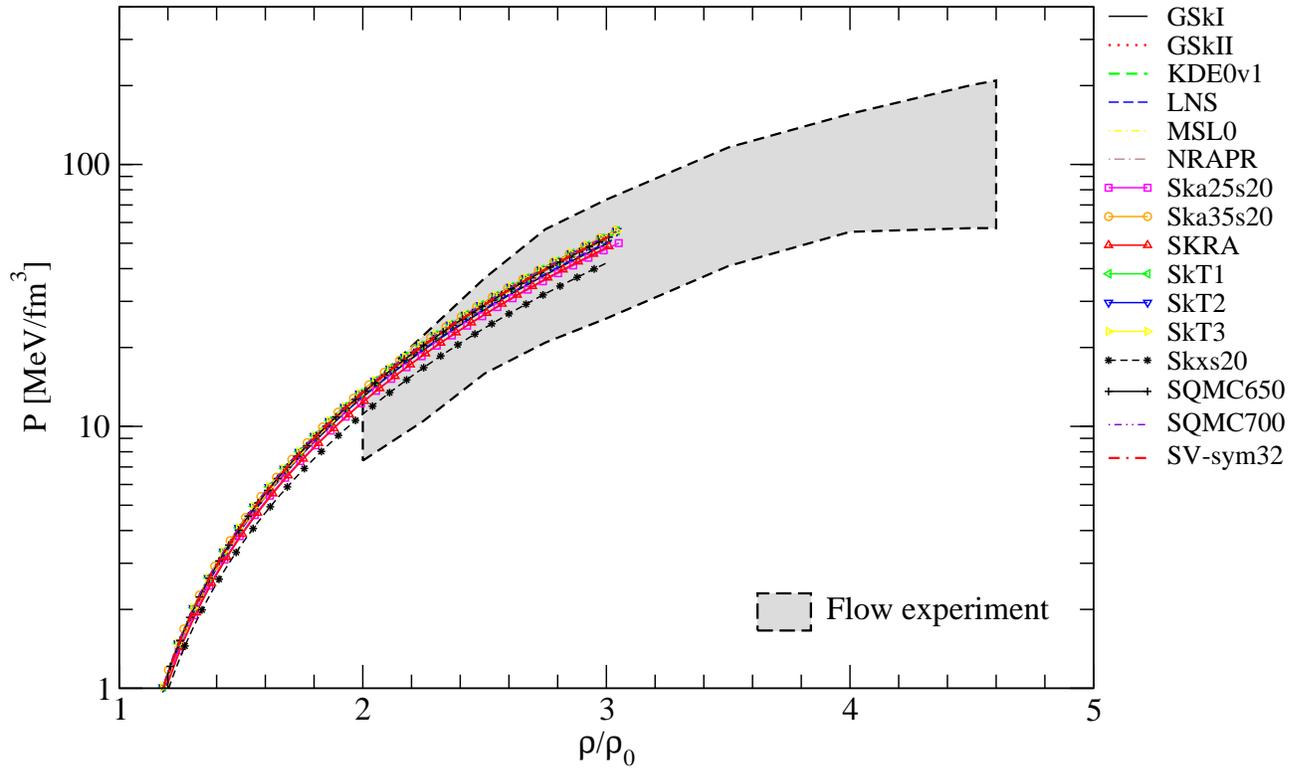}
\caption{(color online) Constraint {\bf{SM3}}: Pressure in SNM as a function of 
density up to 3$\frac{\rho}{\rho_{\rm o}}$ as predicted by consistent Skyrme
parameterizations. The shaded area in the region 2 $< \frac{\rho}{\rho_{\rm o}}
<$ 4.6 is taken from Ref.~\cite{danielewicz2002}.}
\label{fig:sm3}
\end{center}
\end{figure}

\begin{figure}
\begin{center}
\includegraphics[scale=0.6]{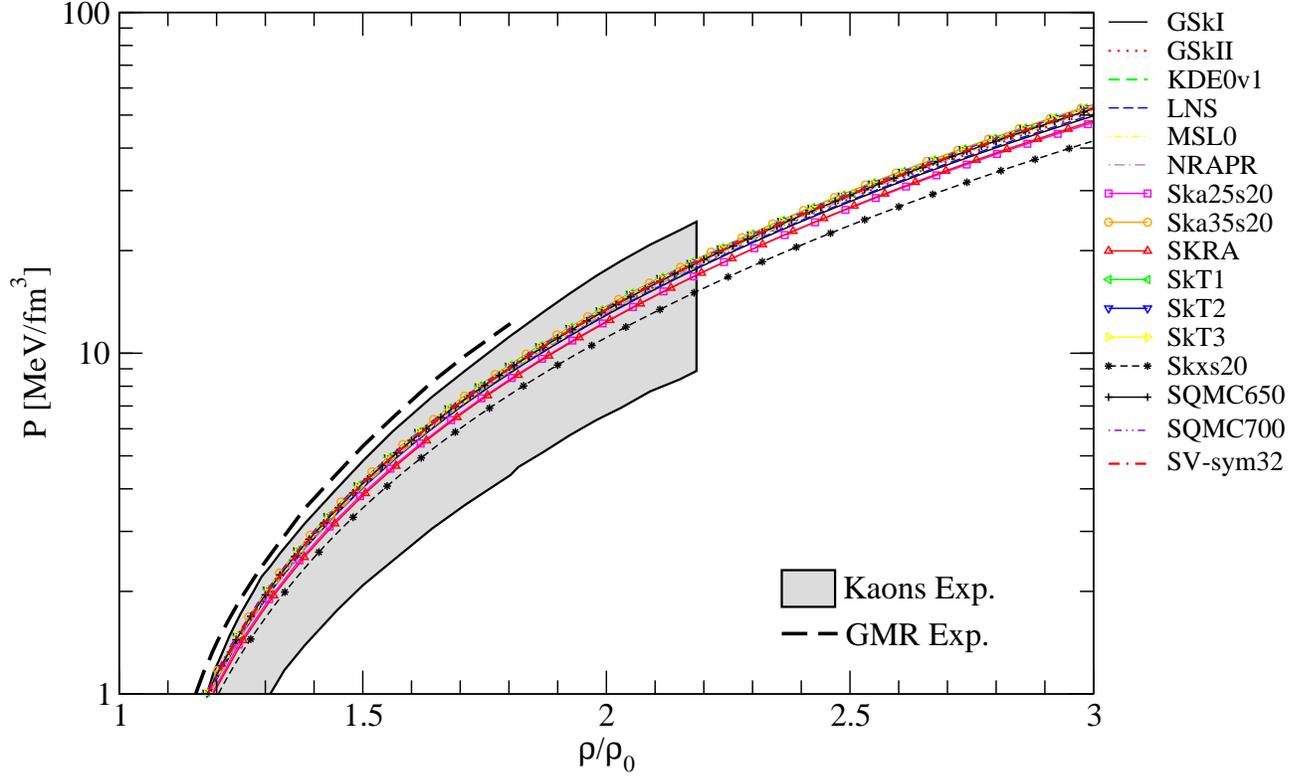}
\caption{(color online) Constraint {\bf{SM4}}:  Pressure in symmetric nuclear 
matter as a function of density in the region 1.2 $<\frac{\rho}{\rho_{\rm o}}
<$ 2.2. The shaded area represents an educated guess discussed in
Ref.~\cite{lynch2009} around the pressure-density relation available from kaon
production data \cite{fuchs2006}. Dashed line extrapolates the pressures
consistent with GMR data to higher densities in the region
1.2~$<~\frac{\rho}{\rho_{\rm o}}~<$~1.7.}  
\label{fig:sm4}
\end{center}
\end{figure}

\begin{figure}
\begin{center}
\includegraphics[scale=0.6]{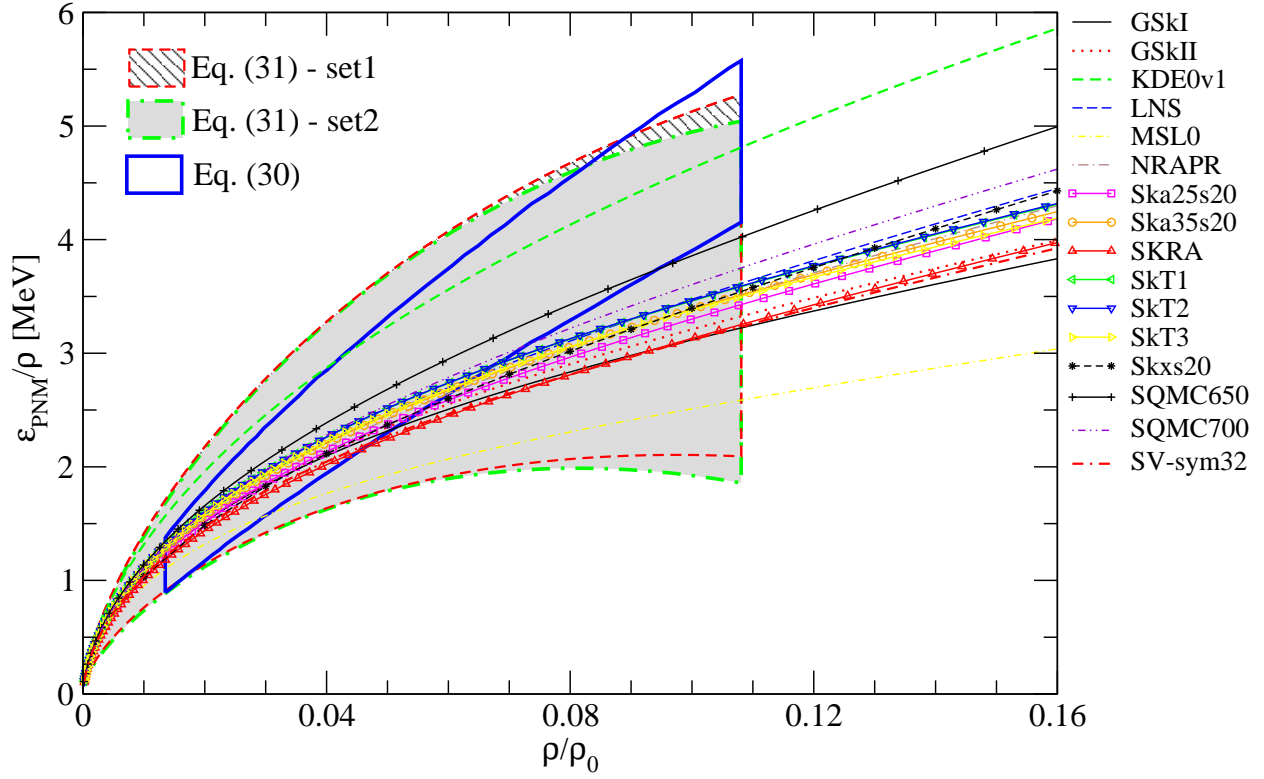}
\caption{(color on line) Constraint {\bf{PNM1}}: Energy per particle in PNM as 
a function of density. The grey bands were based on the Ref.~\cite{epelbaum2009}
for two different parameterizations (dashed red and dashed-dot green lines). The
region inside the solid blue line illustrates the universal limit constraint
given by Eq.~(\ref{unitarity}) \cite{schwenk2005}. For more
explanation see text.}
\label{fig:pnm1}
\end{center}
\end{figure}

\begin{figure}[!htb]
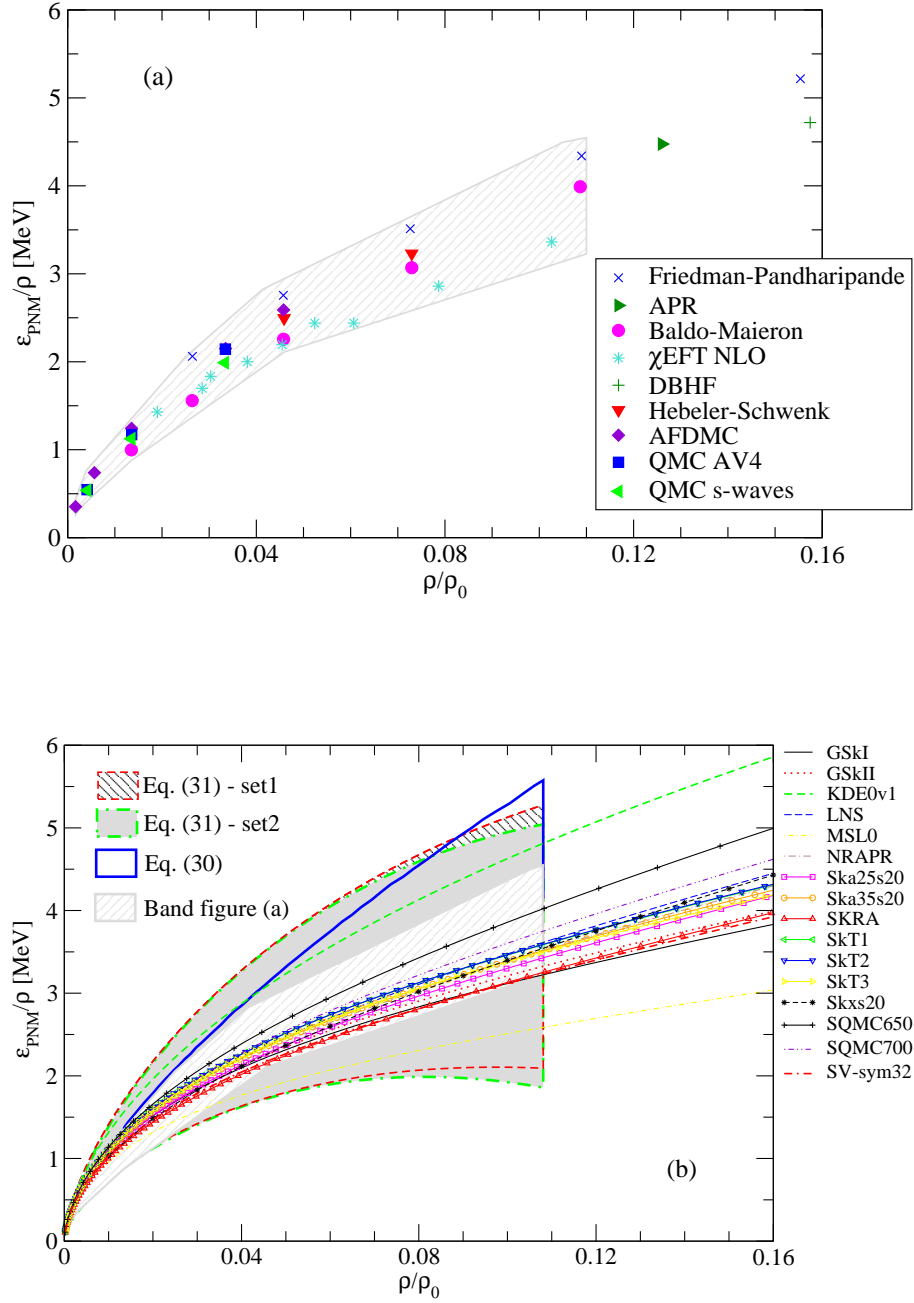

\begin{center}
\includegraphics[width=12cm]{pnm1starpt_0124.eps}\\
\vspace{50pt}
\includegraphics[width=12cm]{pnm1star_0125.eps} \\
\caption{(color online) EoS of low density pure neutron matter: (a) band defined
by results of theoretical calculations summarized in Ref.~\cite{gezerlis2010}.
See the reference for explanation of the legend. (b) the same as in
Fig.~\ref{fig:pnm1}, but with the additional band (a) included.}
\label{fig:pnm1star}
\end{center}
\end{figure}

\begin{figure}
\begin{center}
\includegraphics[scale=0.6]{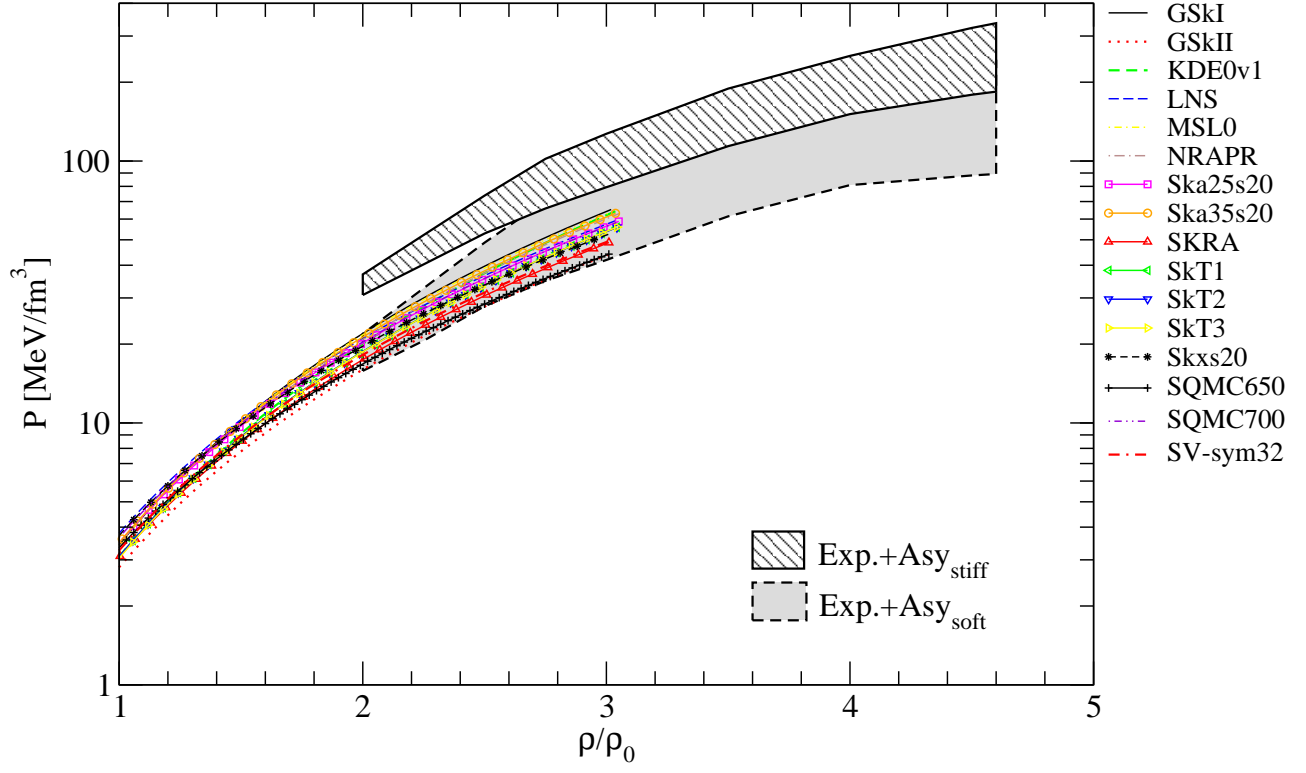}
\caption{(color online). Constraint {\bf{PNM2}}: Pressure in the PNM as a 
function of density as calculated by consistent Skyrme parameterizations up to 
3$\frac{\rho}{\rho_{\rm o}}$. The bands are in the region
2~$<~\frac{\rho}{\rho_{\rm o}}~<$~4.6. For detailed explanation see
Ref.~\cite{danielewicz2003}.}
\label{fig:pnm2}
\end{center}
\end{figure}

The values of all relevant parameters of the CSkP are given in 
Table~\ref{tab:par}. Table~\ref{tab:18} lists all the numerical properties of
nuclear matter at saturation as calculated by the CSkP. The compliance of the
CSkP with graphical constraints SM3, SM4, PNM1, PNM2 is illustrated  in
Figs.~\ref{fig:sm3}--\ref{fig:pnm2}. 

The range of calculated values from all the CSkP is compared with the range
of each constraint in Table~\ref{tab:listcon}, where,  it is interesting to
note, they often fall within a band much narrower than the imposed constraint.
Figs.~\ref{fig:sm3}--\ref{fig:pnm2} show generally the same behaviour for
constraints defined by a function. Two exceptions concern constraint PNM1
(Fig.~\ref{fig:pnm1}) where a narrow band is predicted by all models except
KDE0v1 (high) and MSL0 (low), and MIX5 (Table~\ref{tab:18}), for which all
results are in the upper half of the range, although not closely clustered.

As shown in Fig.~\ref{fig:pnm1star}b, where the new band from
Fig.~\ref{fig:pnm1star}a was added a to the bands and curves already presented 
in Fig.~\ref{fig:pnm1}, two models, KDE0v1 and MSL0 do not satisfy the more
stringent theoretical constraint. However, as we consider PNM1 a valid
constraint in the context of this paper, we keep the KDE0v1 and MSL0
parameterizations in the CSkP list. This issue will be revisited in future.

The relationship between $L$, $\mathcal{S}_{\rm o}$=$J$, and pressure in pure 
neutron matter P$_{\rm o}$ has been examined by Tsang {\it et al.}
\cite{tsang2009}. Based results of mass measurements, HIC, PDR and IAS, they 
produced a composite constraint on these variables which is an extended
variation of our constraint MIX5. Fig.~\ref{fig:tsang}, taken from
(\cite{tsang2012} and references therein), is the latest version of their Fig.~3
in \cite{tsang2009}. Predictions of the CSkP of this relationship all fall
within the blue dashed rectangle, overlapping with all the constraint but
showing no compatibility with the IAS analysis.
 
\section{\label{sec:micro} Microscopic and observational constraints}
In addition to the eleven macroscopic constraints considered in previous 
sections, we introduce some additional, more microscopic constraints and
constraints based on observation of neutron stars and apply them only to already
chosen CSkP. We find that these constraints significantly reduce further the
number of the CSkP, eliminating GSkI, GSkII, MSL0, Ska25s20, Ska35s20, the SkT
group, Skxs20, SQMC650, and SV-sym32 from the CSkP list, as discussed in detail
below.
 
\subsection{\label{sub:effm}The effective mass}
In non-relativistic models of the motion of a nucleon with mass $M$ in 
homogeneous nuclear matter, the nuclear potential V(k) is momentum dependent.
The concept of the effective mass $M^*$, originally developed by Brueckner
\cite{brueckner1955}, leads to an equivalent description of the motion in which
the nuclear potential V(k=0) is constant but the nucleon mass has been modified.
It has been established that the $M^*$ is lower than the $M$ for all potentials
for which the low $k$ expansion V(k)=V(0)+bk$^2$+$\cdots$ (where b is a
constant) is valid.
    
This simple formalism can be extended also to momentum and density dependent 
potentials, such as the Skyrme potential. The nucleon isoscalar effective mass 
$M^*_s$ at saturation density can be defined \cite{klupfel2009} as 
\begin{equation}
\frac{\hbar^2}{2M_s^*}=\frac{\hbar^2}{2M}+\frac{\partial E}{\partial\tau}
\bigg |_{\rho_{\rm o}}
\end{equation}
where $E$ is the energy per particle, Eq.~(\ref{densityenergy}),
which leads to \cite{chabanat1997}
\begin{equation}
m_s^*=\frac{M_s^*}{M}=(1+\frac{M}{8\hbar^2}\rho_{\rm o}\Theta_{\rm s})^{-1},
\label{effms}
\end{equation}
\noindent 
in terms of the Skyrme parameters. 

The isovector effective mass $M^*_v$ is given as 
\begin{equation}
m_v^*=\frac{M_v^*}{M}=\frac{1}{1+\kappa}=\left(1+\frac{M}{4\hbar^2}\rho_{\rm
o}\Theta_{\rm v}\right)^{-1},
\label{effmv}
\end{equation}
where $\kappa$ is the enhancement factor of the Thomas-Reiche-Kuhn
sum rule \cite{ring} 
\begin{equation}
\kappa=\frac{2M}{\hbar^2}\rho_{\rm o}\frac{\partial}{\partial (\tau_n - \tau_p)}
\frac{\partial}{\partial (\rho_n - \rho_p)}E\bigg |_{\rho_{\rm o}}
\end{equation}
(notice the typographic error in \cite{klupfel2009} Eq.~(14)
\cite{reinhard2012}). In the above equations $\tau$,$\tau_p$ and $\tau_n$ are
the total, proton and neutron kinetic energy densities and $\Theta_{\rm s}$ and
$\Theta_{\rm v}$ are defined in Eq.~(\ref{theta}).

In ANM, consisting of unequal amount of neutrons and protons, the nucleon 
effective mass can be written in terms of $\Theta_{\rm s}$ and $\Theta_{\rm v}$ 
as \cite{tondeur1984,farine2001,cao2006}
\begin{equation}
m_q^*=\frac{M_q^*}{M}=\left(1+\frac{M}{8\hbar^2}\rho\Theta_{\rm s}-\frac{M}{
8\hbar^2 }q(2\Theta_{\rm v}-\Theta_{\rm s})\beta\rho\right)^{-1},
\label{aeffm}
\end{equation}
where $\beta$ is the asymmetry parameter ($\rho_n$ - $\rho_p$)/$\rho$  and 
$q$ = 1 (-1) for neutrons (protons).

For non-standard parameterizations the Eq.~(\ref{aeffm}) becomes
\begin{equation}
m_q^* = \left[1+\frac{M}{8\hbar^2}\rho\Theta_{\rm
s}^\prime-\frac{M}{8\hbar^2} q\left(2\Theta_{\rm v}-\Theta_{\rm s}
-t_4(1+2x_4)\rho^\beta + t_5(1+2x_5)\rho^\alpha\right)\beta\rho\right]^{-1}.
\label{aeffmns}
\end{equation}
For such a case, $m^*_s$ and $m^*_v$ are obtained by making $\Theta_{\rm s}
\rightarrow \Theta_{\rm s}^\prime$ and $\Theta_{\rm v} \rightarrow \Theta_{\rm
v}^\prime$ in Eqs.~(\ref{effms}) and (\ref{effmv}), where $\Theta_{\rm
s}^\prime$ and $\Theta_{\rm v}^\prime$ are defined in Eq.~(\ref{thetanp}).

Constraints on the  $m^*_s$ and  $m^*_v$ at saturation density can be derived 
from experimental peak frequencies of giant resonances in finite nuclei
(\cite{klupfel2009,expgr} and references therein). $m^*_s$ is solely related
to giant quadrupole resonance (GQR). Kl\"upfel {\it et al.} \cite{klupfel2009}
deduced an optimum value of $m_s^*$=0.9 from the GQR in $^{\rm
  208}$Pb, close to the 
estimate of Bohias {\it et al.} \cite{bohigas1979} $m_s^*$=0.8. $m^*_v$ is
constrained from the giant dipole resonance (GDR) which is sensitive to two
nuclear matter variables, the symmetry energy and the enhancement factor
$\kappa$. Kl\"upfel {\it et al.}, taking $\kappa$=0.4, obtained $m_v^*$=0.7, the
same as the value extracted in \cite{krivine1980}. However, these values, 
based on the GDR in $^{\rm 208}$Pb, are not quite consistent with data of the
GDR in $^{\rm 16}$O \cite{erler2010}. The current conclusion is that 
experimental GDR data on light and heavy nuclei cannot be satisfied
simultaneously with the present form of the static HF functional. Additional
work is needed in both theory and experiment. 

The effective mass scales the level density $g$ of single-particle (s.p.) 
levels in the vicinity of the Fermi surface
$g(\epsilon_F)\rightarrow\frac{M}{M^*}$g$(\epsilon_F)$ (see \cite{satula2008}
and references therein). This simple scaling, valid in infinite matter, holds
for level density of deep s.p. states in finite nuclei, but breaks down for s.p.
states close to the Fermi level. The origin of the change in level density is
usually seen in coupling between HF s.p. modes and surface-vibration (beyond HF)
RPA modes (see e.g. \cite{bernard1980}). If the coupling is taken into account,
the nuclear matter effective mass scaling can be recovered. Such calculation is
however rather complicated and the simplest way to fit experimental s.p. level
density is to take $M^*$ being state dependent and equal to $M$ at the Fermi
surface \cite{farine2001}. Such an approach is also necessary in fitting atomic
mass data with conventional Skyrme forces, where a high precision fit of masses
of open shell nuclei is not possible without a correct spacing of s.p. states
close to the Fermi surface. However, the choice of $M^*$=$M$ is inconsistent,
for example, in the context of formation of nuclear matter from isolated nuclei
in neutron stars or core-collapse supernova matter. One possible remedy is to
construct an extended Skyrme force \cite{farine2001}. More recently, Satula
{\it et al.} \cite{satula2008}  studied the problem of the effective mass
scaling within the Skyrme energy density functional (EDF) method. They concluded
that more detailed modelling of the two-body spin-orbit and tensor interaction
strength reinstates the conventional $m^*=M^*/M$ scaling and removes the
inconsistency in the effective mass scaling of s.p. level densities in nuclear
matter and finite nuclei, caused mainly by fitting strategies of the Skyrme
parameters to incomplete Skyrme functional.  

Based on the theoretical concept of the effective mass and the experimental data 
on GQR and GDR we find a strong enough reason to eliminate all CSkP
parameterizations with $M^*$=$M$. This choice is not valid in nuclear matter
(the prime concern of this work) and has only an auxiliary character in finite
nuclei which is likely to be improved upon.  The constraint on $m^*_s$ alone
would eliminate the Ska and SkT family forces from the list of CSkP, thereby
reducing their number to eleven. For the remaining, $m^*_v$ at saturation
density is calculated in the range 0.603 - 0.930. Due to the weak nature of this
constraint, deduced from experimental GDR, we do not feel there is a strong
enough reason to do any further elimination.

Next, we examine the density dependence of the effective neutron 
(Fig.~\ref{fig:effmas}, left panel) and proton (Fig.~\ref{fig:effmas}, right
panel) mass, $m^*_n$ and $m^*_p$, in BEM. These are very important in
modelling cold neutron stars. It can be seen that the set SV-sym32 yields  the
$m^*_n$ close to one at very low density, raising rapidly with increasing
density up to about 1.7 at 3$\rho_{\rm o}$. Such behaviour is not physical and
is a reason for elimination of SV-sym32 from CSkP. We note that some other
members of the SV family \cite{klupfel2009} SV-sym28, -sym34, -K218, -K226,
-K241, -bas, -kap60 and -mas10 also exhibit the same feature. They all pass
through a singularity \cite{stone2003} at densities $\sim$5$\rho_{\rm o}$ and
higher.  

\subsection{\label{sub:landau}Landau parameters}
As an alternative to the Hartree-Fock approach to the properties of nuclear 
matter, the formalism of Landau theory of a normal liquid has been used. In this
approach, the bulk properties of nuclear matter are written in terms of a
two-body interaction expressed as a functional second derivative of the energy
per particle with respect to occupational numbers at the Fermi surface. This has
the form
\cite{backman1973,backman1975}.
\begin{equation}
V_{i,j} = \delta(r_i-r_j)N^{\rm -1}_0\sum_L[F_L+F^\prime_L(\tau_i.\tau_j)+
G_L(\sigma_i.\sigma_j)+G^\prime_L(\tau_i.\tau_j)(\sigma_i.\sigma_j)]
P_L(cos\theta).
\end{equation}
The  number of states per unit energy per unit volume at the Fermi surface in 
symmetric matter is $N_0=\frac{2M^*}{\hbar^2}\frac{k_F}{\pi^2}$ where $k_F$ is
the Fermi momentum.  In pure neutron matter this quantity halves. $\theta$
is the angle between the momenta of the interacting particles (holes). 
The sum is over angular momentum L; for the Skyrme interaction, L=0 and 1 as it
contains only S and P wave contributions. The dimensionless parameters $F$ and
$F^\prime$ are directly related to quantities describing nuclear matter such as
effective mass, incompressibility, symmetry energy, and the speed of sound
through relationships \cite{liu1991,margueron2002}
\begin{eqnarray}
m^*_s&=&1+\frac{1}{3}F_1, \label{landaum}\\ 
K&=& 3\frac{\hbar^2 k_F^2}{M^*_s}(1+F_0), \label{landauk} \\
\mathcal{S}&=& \frac{\hbar^2k_F^2}{6 M^*_s}(1+F^\prime_0),\label{landaus}
\qquad\mbox{and}\\
v_s&=&\frac{\hbar^2 k_F^2}{3 M}\frac{1+F_0}{1+1/3F_1}\label{landauv}.
% \label{landau}
\end{eqnarray}
The Landau parameters $G$ and $G^\prime$ determine, to leading order,
properties of nuclear matter in the spin and spin-isospin channels.  We note
that only six, out of the eight Landau parameters in SNM, are independent,
because of two Pauli principle sum rules \cite{friman1979} and conventionally
$F_1^\prime$ and $G_1^\prime$ are expressed as a function of the other six. In
PNM, with no isospin degrees of freedom, only 4 parameters, $F_0$, $G_0$,  $F_1$
and $G_1$ are non-zero.

It can be established that stability demands each of $F_L, F^\prime_L, G_L, 
G_L^\prime$, to be greater than -(2L+1) \cite{backman1975}, i.e., L=0 parameters
must be greater than -1 and L=1 terms greater than -3 (see
Figs.~\ref{fig:landau1} and \ref{fig:landau2}). The most obvious justification
for these conditions is the requirement that incompressibility, symmetry energy
(for stable HF solution for symmetric nuclear matter), and speed of sound be
positive (see Eqs. (\ref{landauk}), (\ref{landaus}), and (\ref{landauv})).

Exact relation between parameters of the Skyrme interaction and Landau 
parameters can be derived (see, e.g. \cite{liu1991,margueron2002}). Beside
the expressions (\ref{landaum}) -- (\ref{landauv}) for $F_0$, $F_1$, and
$F^\prime_0$, we also present the remaining Landau parameters, $G$ and
$G^\prime$ given by
\begin{eqnarray}
G_0 &=& N_0\left[\frac{t_0}{4}(2x_0-1) + \frac{1}{24}
\sum_{i=1}^3t_{3i}(x_{3i}-1)\rho^\sigma+ \frac{t_1}{8}(2x_1-1)k_F^2 +
\frac{t_2}{8}(2x_2+1)k_F^2\right. \nonumber \\
&+& \left.\frac{t_4}{8}(2x_4-1)k_F^2\rho^\delta +
\frac{t_5}{8}(2x_5+1)k_F^2\rho^\gamma\right]\equiv G_0^{\rm SNM},\\
G_0^\prime &=&  N_0\left[-\frac{t_0}{4} - \frac{1}{24}
\sum_{i=1}^3t_{3i}\rho^\sigma - \frac{t_1}{8}(2x_1+1)k_F^2 +
\frac{t_2}{8}(2x_2+1)k_F^2\right. \nonumber \\
&-& \left.\frac{t_4}{8}(2x_4+1)k_F^2\rho^\delta +
\frac{t_5}{8}(2x_5+1)k_F^2\rho^\gamma\right]\equiv G_0^{\prime\rm SNM}, \\
G_1 &=& -N_0k_F^2\left[\frac{t_1}{8}(2x_1-1)+\frac{t_2}{8}(2x_2+1)
+ \frac{t_4}{8}(2x_4-1)\rho^\delta + \frac{t_5}{8}(2x_5+1)\rho^\gamma\right]
\nonumber \\
&\equiv& G_1^{\rm SNM},
\end{eqnarray}
for SNM \cite{chamel2009}, and
\begin{eqnarray}
G_0 &=& G_0^{\rm SNM} + G_0^{\prime\rm SNM}, \qquad\mbox{and} \\
G_1 &=& G_1^{\rm SNM} +
\frac{N_0k_F^2}{8}\left(t_1 - t_2 + t_4\rho^\delta-t_5\rho^\gamma\right)
\end{eqnarray}
for PNM \cite{goriely2010}, also valid for the non-standard Skyrme models.

Fig.~\ref{fig:landau1} (for SNM) and Fig.~\ref{fig:landau2} (for PNM) show the 
results of the variation of the Landau parameters with density for the CSkP
sets, after application of the constraint related to effective mass. It can be
seen, at densities below about $\sim$0.1 fm$^{\rm -3}$, that all CSkP sets
predict negative incompressibility in SNM. This feature, referred to as a
spinodal instability, should be seen as a shortcoming but also a realistic
consequence of the strong correlations between nucleons which, at low density,
cause them to form a kind of gas-liquid separation. This instability was
observed experimentally in heavy-ion collisions \cite{suraud1989} at critical
densities about half to two-thirds of nuclear saturation density. No such
transition is predicted in PNM (Fig.~\ref{fig:landau2}). This instability in
symmetric and BEM matter has been consistently predicted by a variety of
non-relativistic HF and RMF models \cite{ducoin2008}.

The density dependence of the parameter $G_0$ can be used as an indicator of a 
breakdown of spin symmetry, i.e., a transition to a spin-ordered (ferromagnetic)
phase in SNM as well as in PNM. Such a transition could have important
consequences for the evolution of the proton-neutron star in core-collapse
supernova and neutrino transport inside the star \cite{reddy1999}, but is not as
yet constrained  experimentally. Theoretical studies of the spin-ordered phase
yield rather contradictory results. Skyrme interactions predict such a
transition at low densities (below 3$\rho_{\rm o}$) in PNM, SNM and ANM
\cite{viduarre1984, reddy1999, isayev2003, isayev2004, rios2005}. Relativistic
DBHF calculations with an effective Lagrangian also predict a transition to a
spin-ordered phase at several times $\rho_{\rm o}$ (see \cite{bernardos1995} and
references therein). On the contrary, realistic NN interactions suppress such a
transition up to high densities in BHF models \cite{vidana2002, vidana2002a},
the AFDMC method \cite{fantoni2001}, and lowest order constrained variational
method \cite{bordbar2008}.

As shown in Fig.~\ref{fig:landau1}, the transition to spin-ordered matter is 
predicted in SNM at densities below $\rho_{\rm o}$ for the GSkI and GSkII, and 
at around 1.5 $\rho_{\rm o}$ for MSL0, Skxs20 and SQMC650 parameterizations. In
PNM, five parameterizations predict the transition at densities below and around
$\rho_{\rm o}$, GSkI, GSkII, MSL0, Skxs20 and SQMC650 and two at around 1.5
$\rho_{\rm o}$, SKRA and SQMC700.

These features are not realistic and we eliminate GSkI, GSkII, MSL0, Skxs20 and
SQMC650 in their present form from the CSkP list as they did not satisfy both,
the SNM and PNM constraint. However, as demonstrated by Margueron and Sagawa
\cite{margueron2009}, the spin-spin and spin-isospin instabilities can be removed
if additional density dependent terms, affecting only the spin and spin-isospin
channels, are included in the standard Skyrme Hamiltonian. The contribution of
the new terms to the mean field is zero for spin-saturated systems. Consequently
the properties of the original Skyrme interaction are not changed in this case.
However, in nuclear matter four new parameters have to be adjusted to values of
Landau parameters at saturation density extracted from the G-matrix. This
procedure is rather involved and is beyond the scope of the present work. We
suggest that parameterizations SKRA and SQMC700 would be the best candidates for
application of treatment \cite{margueron2009} in future.   

We note that the density dependence of the parameter $G_0^\prime$, indicates a
spin-isospin instability if $G_0^\prime$ falls below -1. Such instability has 
been interpreted as the appearance of a pion condensate \cite{akmal1997,
migdal1978}. Of the remaining parameterizations, only KDE0v1 and LNS predicts
such transition in SNM below 3$\rho_{\rm o}$. All the other Landau parameters 
are within the natural constraints.
 
\subsection{\label{sub:ddsym}Density dependence of the symmetry energy}
One rather surprising result, which came out of our analysis, is that the CSkP 
exhibit a growing spread in density dependence of the symmetry energy beyond
about twice the nuclear saturation density. This feature is illustrated in
Fig.~\ref{fig:esym} and Table~\ref{tab:3rho} and suggests that constraining the
derivatives of the symmetry energy at the saturation point is not sufficient for
controlling the slope of $\mathcal{S}$($\rho$) at higher densities. Clearly,
more experimental data is needed to constrain the Skyrme interaction at
super-saturation densities. 

It turns out that, considering the symmetry energy being the difference between 
the energy per particle in pure neutron and symmetric matter (to the first
order), the factor which mainly determines the behavior of the symmetry energy
with increasing density is the pure neutron matter EoS. In Fig.~\ref{fig:sym3f}
we see energy per particle in PNM and SNM as a function of density as calculated
with Skxs20, QMC700 and GSkII parameterizations. These forces were chosen as
they correspond to the top, middle and bottom curves in Fig.~\ref{fig:esym}. We
see clearly that the energy per particle for SNM are rather similar in all
three panels, but for PNM it exhibits systematically a different pattern. In a
sense this is not surprising. Skyrme parameterizations are usually fitted to
properties of nuclei with either $N=Z$ or a low value of isospin. The EoS for
PNM is well constrained at low densities; at super-saturation densities we have
to rely on theoretical models or seek indirect evidence from astrophysical
extrapolations, e.g. to neutron stars. So constraining the PNM EoS by study of
very neutron rich heavy nuclei should be desirable.

\subsection{\label{sub:hmns}High-mass cold neutron stars}

One possibility is to use the Skyrme EoS in cold neutron star models up to
3$\rho_{\rm o}$. Here the Skyrme interaction is applied to $n+p+e+\mu$ BEM
rather then symmetric or pure neutron matter. There are no constraints available
from terrestrial experiments  at present, as the heavy-ion reactions are too
fast to build equilibrium conditions with respect to weak interactions such as
$p+e^- \leftrightarrow p + \mu^-$. 

The findings of Ref.~\cite{stone2003} indicate that, if one accepts
validity of the  Skyrme interaction at densities up to 10 times
nuclear saturation density, only parameterizations predicting growing
(or monotonously slowly decreasing - for details see Ref.~\cite{stone2003})
symmetry energy with increasing density can be used to generate stable neutron
star models with mass and radius consistent with currently available
observational data \cite{lattimer2007}. We observe the same phenomenon here for
selected CSkP as illustrated in the left panel of Fig.~\ref{fig:mr}. Here the
mass-radius relation for neutron star models was calculated using a BEM Skyrme
EoS; since only a part of the neutron star is in the BEM phase, we used the
Baym-Pethick-Sutherland EoS (BPS) \cite{baym1971} at lower densities matching
the Skyrme based part at $\rho \sim$ 0.1 fm$^{\rm -3}$ and going down to  $\rho
\sim$ 6.0$\times$10$^{\rm -12}$ fm$^{\rm -3}$.  The observed spread in the
maximum mass models is not unexpected; it is related to the different
extrapolation properties of the Skyrme interaction to densities well beyond the
validity of the Skyrme model. A very recent observational finding of a massive
neutron star \cite{demorest2010} with M$\rm_g$=1.97$\pm$0.04 M$_\odot$ and the
central density less than $\approx$ 10$\rho_{\rm o}$ provides a strong
constraint on EoS of BEM. This constraint would certainly eliminate SKRA,
SQMC700, and LNS EoS which predict lower maximum mass models. Moreover, the
central densities  of \textit{all} maximum mass neutron star models predicted by
CSkP, including those within the window set by the results of Demorest {\it et
al.} for gravitational mass, predict higher central densities than is allowed
(in the region for 11-13$\rho_{\rm o}$). We thus conclude that extrapolation of
the Skyrme model beyond its validity range to high densities does not predict
cold neutron stars in agreement with the recent observation. We do not eliminate
any Skyrme interaction from the CSkP list on the basis of this constraint
because it requires extension of the Skyrme model outside its validity range.
     
An alternative is to use the Skyrme interaction within its validity range 
argued in this paper (up to about 3$\rho_{\rm o}$) to construct the EoS of BEM
and match it to an established high density EoS, as well as with BPS EoS at
lower densities. Such an EoS was usually thought to be the Bethe-Johnson EoS
\cite{bethe} based of the Reid potential and including hyperons in the composite
matter at high densities. However, the Demorest {\it et al.} observation rules
out this EoS as it predicts maximum mass of the neutron star models to be only
$\approx$ 1.85 M$_\odot$ with central density $\approx$ 12$\rho_{\rm o}$.
Therefore we use the Full-Quark-Meson-Coupling model (FQMC) \cite{stone2007a}
which includes full baryon octet in the high density matter and  provides high
mass neutron star models in agreement with observation,  as shown in the right
panel of Fig.~\ref{fig:mr}. The maximum mass is clearly determined by
the FQMC model and only the variations in radii of smaller neutron star models 
with smaller central densities are due to different Skyrme EoS.
Observational data on neutron star radii are very poor at present. However, 
new observational techniques are being developed and radii may be known within a
few percent in the near future. They could be then used as a useful constraint
on the Skyrme parameterizations performance in a high density neutron rich
environment. 

\subsection{\label{sub:lmns}Low-mass cold neutron star}
Observation of the double pulsar J0737-3039 provides another, stringent, 
constraint on the neutron star EoS and its interpretation by
Podsiadlowski {\it et al.} \cite{podsi2005}, and hence on the
effective nucleon-nucleon interaction in stellar matter. This
constraint is important, as the central density of pulsar B is only
2-3$\rho_{\rm o}$ and thus the Skyrme interaction is expected to be 
fully applicable.

The constraint concerns the ratio of the gravitational mass of pulsar B to 
its baryonic mass. The gravitational mass of Pulsar B is very precisely known
M$_{\rm g}$=1.249$\pm$0.001 M$_\odot$. Estimates of the baryonic mass depend
upon its detailed mode of formation.  If the pulsar B was formed from a white
dwarf with an O-Ne-Mg core in an electron-capture supernova, assuming no or
negligible loss of baryonic mass during the collapse, the newly born neutron
star will have the same baryonic mass as the precollapse core of the progenitor
star. As modelled by Podsiadlowski {\it et al.} \cite{podsi2005}, the baryonic
mass of pulsar B is then between 1.366 - 1.375 M$_\odot$. The range reflects
uncertainties in modelling of the progenitor core such as electron-capture
rates, nuclear network calculations and Coulomb and general relativity
corrections. Another simulation of the same process by Kitaura {\it et al.}
\cite{kitaura2006} gave M$_\odot$ 1.360$\pm$0.002 M$_\odot$ where the range of the
result was mainly due to uncertainty in the EoS and the estimated small mass
loss during the collapse. 

For any neutron star matter EoS the relation between the gravitational and 
baryonic mass can be calculated. Fig.~\ref{fig:pulb} shows the results for the
remaining five CSkP to be checked against narrow windows given by the two models
of pulsar B. All of the CSkP but NRAPR, which is slightly shifted, agree 
remarkably well with the result of Kitaura {\it et al.} \cite{kitaura2006} thus
supporting the concept of some baryonic mass loss during the collapse.

\section{\label{sec:disc}Discussion and summary}

In embarking on this project we hoped the present work might lead to better 
understanding of how a universal Skyrme force might best be achieved. In this
section we discuss the degree to which this hope has been accomplished.
A distinction was made between standard and non-standard forms of the 
Skyrme interaction, the latter including additional terms added relatively
recently. The overall results are discussed below for the two forms.  

\subsection{\label{sub:stand}Standard parameterizations}

Only six of the over 224 standard sets satisfied all the constraints. 
Examination of Table~\ref{tab:const} reveals that some families of sets show
systematic patterns of failure, as discussed briefly below. Other
parameterizations, among them several commonly used for years in modelling
finite nuclei, for example the SIII, SkM*, SkP, SGI, SGII and the SkI and SKX
families, each fail different constraints with no apparent pattern. 

This very variable performance makes it rather difficult to identify a particular 
term or terms in the Skyrme energy functional as responsible. Among the more
systematic inconsistencies, BSk1-17 forces, with the exception of BSk14, do not
satisfy constraints PNM2, MIX2 and MIX3, a feature shared with most of their 
predecessors, the SkSC and MSk families. They predict too low pressure in PNM as
a function of increasing super-saturation density. They underpredict both the
symmetry energy and its derivative at the saturation density (see
Table~\ref{tab:237} and Fig.~\ref{fig:tsang}). In addition, the failure of MIX3
indicates too high volume part of the isospin incompressibility, i.e. they
overpredict isospin dependence of the curvature of the EoS of ANM
E($\rho$,$\beta$) at the saturation density $\rho(\beta)$ (see
Eq.(\ref{erhobeta})).   

None of the well known Lyon forces satisfy constraint MIX3, although they all 
pass constraint MIX2 (except for SLy10), similarly to the BSk forces. It is 
interesting to note that SLy230a force \cite{chabanat1997}, especially developed
for modelling of neutron stars, fails constraint PNM1, i.e. does not have the
correct density dependence of pressure at sub-nuclear densities, especially
important in modelling of neutron star crusts. The over-prediction of K$_{\rm
\tau,v}$ is also a feature of recently developed extensive set of new Skyrme
forces \cite{lesinski2007} using a fitting protocol similar to that used for
construction of Lyon forces.
 
Failure of MIX3  may seem to be a minor defect but it is persistent and points 
to the isospin part of the Skyrme force. Isospin dependence of the curvature of
the EoS of ANM plays an important role in modelling of giant resonances and
heavy-ion collisions \cite{bali2008,chen2009} 

The very recently developed modified Skyrme-Like (MSL) model \cite{chen2010}, 
which is expressed in terms of 9 macroscopic observables that are either
constrained experimentally or well known empirically, offers another
non-traditional approach to the  construction of a Skyrme parameterization. It
expresses the standard Skyrme parameters in terms of these observables and
provides a parameterization MSL0 that complies with all but one of the 
constraints studied in this work. It predicts spin instability around nuclear
saturation density, which would be a problem, especially in using this force to
model neutron stars.

The successful CSkP sets, several of which were unfortunately infrequently used,
do not share much common ground. Their individual parameters, listed in
Table~\ref{tab:par}, are too spread to give a useful guidance for construction
of a more general ``consistent'' set. 

In recent years some effort, aimed at giving more physical relevance to 
particular terms in the Skyrme energy functional, has lead to its re-expression
in terms of the coupling constants (some of them density dependent) involving
linear combinations of the individual parameters in Eq.~(\ref{densityenergy}).
Four of these coupling constants, relevant for calculations of nuclear matter,
are
\begin{eqnarray}
C^\rho_{\rm o}&=&\frac{3}{8}t_0+\frac{3}{48}t_3\rho_{\rm o}^\sigma, \nonumber \\
C^\rho_1&=&-\frac{1}{4}t_0(\frac{1}{2}+x_0)-\frac{1}{24}t_3(\frac{1}{2}
+x_3)\rho_{\rm o}^\sigma,\nonumber \\
C^\tau_{\rm o}&=&\frac{3}{16}t_1+\frac{1}{4}t_2(\frac{5}{4}+x_2) \nonumber \\
C^\tau_1&=&-\frac{1}{8}t_1(\frac{1}{2}+x_1)+\frac{1}{8}t_2(\frac{1}{2}+x_2).
\label{constc}
\end{eqnarray}
Numerical values of these coupling constants at saturation density are given 
for CSkP in Table~\ref{tab:comb}. Again, unfortunately, they do not exhibit any
apparent regularity.

Another combination of parameters, involving only $t_1,t_2,x_1,x_2$, has been
introduced \cite{chabanat1997}
\begin{eqnarray}
 \Theta_{\rm s} &=&3t_1+(5+4x_2)t_2, \nonumber \\
 \Theta_{\rm v} &=&t_1(2+x_1)+t_2(2+x_2), \nonumber \\
 \Theta_{\rm sym}&=&3t_1x_1-t_2(4+5x_2), \nonumber \\
 \Theta_{\rm n} &=&t_1(1-x_1)+3t_2(1+x_2).
\label{theta}
\end{eqnarray}
\noindent
$\Theta_{\rm s}$ and $\Theta_{\rm v}$ were used in Sec.~\ref{sub:effm} 
in connection with the effective mass. $\Theta_{\rm sym}$ is used in calculation
of the symmetry energy and $\Theta_{\rm n}$ appears in the expression for
the EoS of pure neutron matter. Numerical values, given in
Table~\ref{tab:comb} for the CSkP, again show a large scatter.  

We are forced to the conclusion that our analysis of the performance of the 
standard parameterizations in nuclear matter does not offer any clear direction
for the development of a unified, generally applicable, Skyrme parameterization.
Indeed, the overwhelming impression, however well intentioned, is that they are
merely empirical attempts to describe nuclear matter related phenomena. We see
the problem as lying first with the lack of a direct connection between the
terms of the Skyrme energy functional and specific physics observables, and
second with the strong correlations between the parameters. 

\subsection{\label{sub:nonst}Non-standard parameterizations}
In addition to the standard definition of the Skyrme energy density functional, 
some recently developed Skyrme models include higher order terms, thus
introducing  additional parameters.  The standard form (for application in
nuclear matter) depends on $9$ parameters,  $t_j$, $x_j$ (with $j=0-3$) and
$\sigma$, whereas the extended Skyrme models considered in this work have
$t_{3i}$ and $\sigma_{3i}$ (with $i=2,3$), $t_4$ and $t_5$ in addition, totaling
15 adjustable parameters.

Out of  all Skyrme models we analysed, $16$ are non-standard, and some of those 
share their origin with the standard ones and are closely related to them. For
example, the BSk family consists of $22$ individual models, $4$ of them
(BSk18-21) being non-standard. The BSk18 model behaves as a standard at SNM,
since  the contribution of the additional terms included adds up to zero. The
extended BSk parameterizations were generated by Goriely {\it et al.} to improve
calculation of nuclear masses at the neutron drip-line for $Z, N \geq 8$ and
$Z\leq 120$. The terms containing  $t_4$ and $t_5$ were introduced to ensure a
better description for homogeneous neutron matter \cite{goriely2010}.
With the same aim, and by adding to the conventional Skyrme forces higher order
density terms in the EoS ($t_{32}$ and $t_4$ contributions
of Eq.~(\ref{densityenergy})), Farine {\it et al.} suggested the SkPS.1 force
\cite{farine2001}. These authors claim that this force fitts well the nuclear
masses of spherical nuclei, and may be useful to describe stelar collapse
processes. Following the same protocol of the ``Saclay-Lyon'' group
(SLy-forces), and by using a better control regarding
the spin-isospin instabilities via Landau parameters, T. Lesinski {\it et al.}
\cite{lesinski2006} have also developed three non-standard forces 
(f$_-$, f$_0$ and f$_+$). They have attempted to constrain the effective neutron
mass in such a way that $m_n^*<m_p^*$, $m_n^*=m_p^*$, and $m_n^*>m_p^*$,
respectively for f$_-$, f$_0$, and f$_+$. By keeping the $t_4$ term in the
EoS, Eq.~(\ref{densityenergy}), and aiming at a good description of excited 
states of finite nuclei, K. Krewald {\it et al.} suggested six new non-standard
Skyrme parameterizations: GS1-6 \cite{Li-Guo-Qiang1991,krewald1977}. Agrawal
{\it et. al.} \cite{agrawal2006}, by exploring the extended density-dependent 
Skyrme effective forces for normal and isospin-rich nuclei for neutron stars,
parameterized two non-standard Skyrme forces (GSkI and GSkII) by adjusting
$t_{32}$ and $t_{33}$. These models were able to fit consistently thirteen
finite nuclei:$^{16}$O, $^{24}$O, $^{14}$Ca, $^{48}$Ca, $^{48}$Ni, $^{56}$Ni,
$^{68}$Ni, $^{78}$Ni, $^{88}$Sr, $^{90}$Zr, $^{100}$Sn, $^{132}$Sn, and 
$^{208}$Pb. The breathing modes for $^{90}$Zr and $^{208}$Pb were also well
described.

From all the above listed Skyrme non-standard forces, only two, namely, GSkI and
GSkII, satisfied the macroscopic constraints but failed the microscopic ones,
namely the value of the Landau parameter $G_{\rm 0}$. Inclusion of any
nonstandard piece in the energy density functional, Eq.~(\ref{densityenergy}),
inevitably affects parameters of the standard part as both contributions have to
be compensated to fit experimental data. Nevertheless, it is instructive to
investigate the non-standard contribution to the energy per particle of
symmetric matter, obtained from Eq.~(\ref{densityenergy}) and shown in
Fig.~\ref{fig:nonstand}. We see that non-standard terms may either increase
repulsion (attraction) by a positive (negative) term in the effective Skyrme
force. It is interesting to notice that both GSkI and GSkII forces receive very
similar large negative contribution from the non-standard terms apparently
needed to compensate repulsion coming from the standard part of the interaction.
However, because of this delicate balance between the standard and non-standard
terms it is difficult to find any general trend. 

As in the case of the standard Skyrme parameterizations, one can also define the
following set of coupling constants \cite{bender2003},
\begin{eqnarray}
C_{\rm o}^{\rho^\prime}&=&\frac{3t_0}{8}+\frac{3}{48}\sum_{i=1}^3t_{3i}\rho_{\rm
o}^{\sigma_i } ,\nonumber \\
C_1^{\rho^\prime}&=&-\frac{t_0}{4}\left(\frac{1}{2}+x_0\right)-\frac{1}{24}
\sum_ { i=1}^3 t_{3i}\rho_{\rm o}^{\sigma_i}\left(\frac{1}{2}+x_{3i}\right),
\nonumber \\ 
C_{\rm o}^{\tau^\prime}&=&\frac{3t_1}{16}+\frac{t_2}{4}\left(\frac{5}{4}
+x_2\right)+\frac{3t_4} {16}\rho_{\rm o}^\delta+\frac{t_5}{4}\left(\frac{5}{4}
+x_5\right)\rho_{\rm o}^\gamma, \nonumber \\ 
C_1^{\tau^\prime}&=&-\frac{t_1}{8}\left(\frac{1}{2}+x_1\right)+\frac{t_2}{8}
\left(\frac{1}{2}+x_2\right)-\frac{t_4}{8}\left(\frac{1}{2}
+x_4\right)\rho_{\rm o}^\delta+\frac{t_5}{8}\left(\frac{1}{2}
+x_5\right)\rho_{\rm o}^\gamma ,
\end{eqnarray}
written as a function of the non-standard parameters. Notice that these
equations are generalized forms of the Eq.~(\ref{constc}). The same occurs for
the quantities \cite{chabanat1997}
\begin{eqnarray}
\Theta_{\rm s}^\prime &=& \Theta_{\rm s} + 3t_4\rho^\delta
+t_5(5+4x_5)\rho^\gamma,
\label{thetasp}\nonumber \\
\Theta_{\rm v}^\prime &=& \Theta_{\rm v} + t_4(2+x4)\rho^\delta
+t_5(2+x_5)\rho^\gamma,
\label{thetavp}\nonumber \\ 
\Theta_{\rm sym}^\prime &=& \Theta_{\rm sym} + 3t_4x_4\rho^\delta -
t_5(4+5x_5)\rho^\gamma \label{thetasymp},\nonumber \\ 
\Theta_{\rm n}^\prime   &=& \Theta_{\rm n} + t_4(1-x_4)\rho^\delta +
3t_5(1+x_5)\rho^\gamma\label{thetanp}.
\end{eqnarray}
All these values are shown in Table~\ref{tab:comb}.

Although it may be useful to extend the Skyrme functional to improve results 
in particular physical situation, it does not seem to be a way forward to
finding a recipe for getting the Skyrme model under control.    

\section{\label{sec:concl}Conclusions}
We have examined the performance of 240 different Skyrme model parameterizations
in nuclear matter by comparing their predictions of behaviour in eleven areas in
which experimentally/empirically derived constraints exist. The chosen
macroscopic constraints cover a wide range of properties related to Symmetric
Nuclear Matter (SM1-SM4), Pure Neutron Matter (PNM1-PNM2) and both SNM and PNM
(MIX1-MIX5). Of the Skyrme models six satisfy all the constraints whilst 66
satisfy all but one. For ten of the 66 the single failure is narrow, less then 
5\%. Including these yield a final list of 16 consistent models, the CSkP set
{\bf GSkI, GSkII, KDE0v1, LNS, MSL0, NRAPR, Ska25s20, Ska35s20, SKRA, Skxs20,
SQMC650, SQMC700, SkT1, SkT2, SkT3, and SV-sym32}. The parameters of these
interactions are summarized in Table~\ref{tab:fit}.  

As an additional step, we considered four microscopic constraints arising from 
giant resonance experiments on nucleon effective mass, Landau parameters in SNM
and PNM, and observation of low-mass neutron stars. With these microscopic
constraints taken into account the successful set reduces to five,  {\bf KDE0v1,
LNS, NRAPR, SKRA, and SQMC700}, the CSkP* set.

A fifth microscopic constraint, maximum mass and the corresponding central 
density of high-mass neutron stars creates a fundamental obstacle to applying
Skyrme (nucleon only) models in neutron star modelling since it requires
extrapolation to densities above the range of validity. None of the CSkP models
produce a maximum mass neutron star model with central density in line with
observation. Thus if this constraint were applied, all CSkP would fail.
 
We were unable to identify any regularities, either in single parameters or in
their combinations, to identify a unique quality of the CSkP* sets. This is
hardly surprising when we consider the number of parameters and their
correlations, and it is tempting to suggest that some of the sets satisfied all
the constraints by a fortunate accident. This looks particularly likely in the
case of the  KDE family. Of very close KDE0 parameter sets (KDE0v0, KDE0v1), 
which fit the same experimental data but differ only by starting conditions for
the simulated-annealing fit procedure, only KDE0v1 passes our constraints. It
may be significant that in this procedure particular attention was paid to the
inclusion of nuclear matter quantities in the fit. Also energies of the giant
monopole resonances were included directly to the fit for the first time.   

The example of KDE0v1 indicates that the inclusion as many constraints as 
possible, both macroscopic and microscopic, in the fitting protocols of the
Skyrme interaction is essential. For example, we believe that the symmetry
energy plays a key role in the behaviour of nuclear matter. Therefore the
correct determination of the PNM EoS is imperative. Whilst there are no direct
experimental or observational data on PNM at super-saturation densities,
promising {\it ab initio} theoretical predictions and indirect experimental data
from cold atoms at sub-saturation densities are becoming available. Any further
development in this area is very desirable.

It may be also revealing that in construction of three out of five CSkP* 
parameterizations,  LNS, NRAPR, and SKRA, the EoS of nuclear matter, used in the
fit, was derived from realistic potentials. Such an approach amplifies the role
of the microscopic physics input in the effective Skyrme interaction model. The
usual practice of including basic properties of nuclear matter only at
saturation density is not sufficient. The {\it density dependence} of these
observables (within the range of applicability of the Skyrme model), which may
considerably influence the fits must be included.

The last parameterization on the CSkP* list, SQMC700, is unique in its 
derivation from the Quark-Meson-Coupling model (FQMC)
\cite{guichon2004,guichon2006,stone2007a}, which includes the full baryon octet
at high densities and is relativistic. The structure of the QMC Hamiltonian and
its Skyrme equivalent (FQMC limit at low densities) are somewhat different,
which shows up most markedly in the difference between the values of the
incompressibility for FQMC EoS \cite{stone2007a} and QMC forces considered in
this paper. Nevertheless, the fact that in FQMC the many-body interaction is
directly related to the response of quark structure to the nuclear environment
may cause the very simple QMC Skyrme parameterization, with $x_1$=$x_2$=$x_3$=0
and $\sigma$=1/6, to be consistent with all the constraints.

It is important to stress that our present work is restricted to examining the 
performance of the Skyrme interaction in nuclear matter related scenarios. It is
generally true that only a limited effort was spent to make the CSkP* perform
well in wide-ranging application in finite nuclei when they were
derived in the original papers. In some sense, this may be a positive factor as 
it seems obvious that Skyrme interactions, constructed with emphasis on nuclear
matter properties, will do better on constraints derived from nuclear matter
then interactions heavily biased towards properties of finite nuclei. As a
follow-up project of this work, the performance of CSkP in  finite nuclei will
be studied in more detail \cite{stevenson2012}. It will be interesting to test
the quality of the Skyrme functional by extending the fitting protocol to 
include not only all the constraints studied in this work but also requirements
based on  the most up-to-date finite nuclei data, including drip-line and
superheavy nuclei. It may turn out that the ambition to fit using such a
detailed protocol may be asking too much of the Skyrme model (with its known
deficiencies \cite{erler2010}) but the attempt may also lead to a
parameterization with increased predictive power. It is our opinion that a
parameterization, successful in reproducing only a selected subset of available
data but failing the rest, does not have credible predictive power and does not
progress the field. 

Results obtained in this paper should be seen as the first step in a global 
effort to find the best possible Skyrme interactions for use in modeling of
nuclear matter. We intend to monitor developments both in keeping up-to-date the
existing constraints and adopting new ones. Progress in {\it ab-initio}
calculations of inhomogeneous neutron matter \cite{gandolfi2011} and chiral
effective field theory \cite{hebeler2010, hebeler2010a} are good examples of
sources of new constraints which will be taken into account in future. New
experimental results on giant resonances, neutron skin in heavy nuclei and heavy
ion collision, as well as new astrophysical observations will further improve
the set of benchmark constraints, which may shed more light on the structure and
applicability of the Skyrme interaction. Should new Skyrme parameter sets appear
in the literature, we intend to catalogue and test them using the most complete
set of constraints available to us.
  
The outcome of our analysis of all, standard and non-standard, parameterizations
does not offer a final solution to the ``Skyrme proliferation'' problem. Neither
does it provide general guidance for construction of more Skyrme parameter
sets. Production of new parameter sets having limited range of application
should not be encouraged. Rather, more emphasis should be put on better
understanding of the existing models, which should be further tested against an
extended number of refined constraints including both, nuclear matter {\it and}
finite nuclei related properties, with equal emphasis. The Skyrme interaction
has played a dominant role in low energy nuclear physics for decades. The
approach suggested can lead to the final judgment whether or not this
interaction includes enough physics not only for a successful interpretation,
but also for a  prediction, of the rich variety of data and observations on
nuclear and astrophysical systems available today and expected in future.

\section*{ACKNOWLEDGEMENTS}

\begin{figure}
\begin{center}
\includegraphics[angle=90,width=14cm]{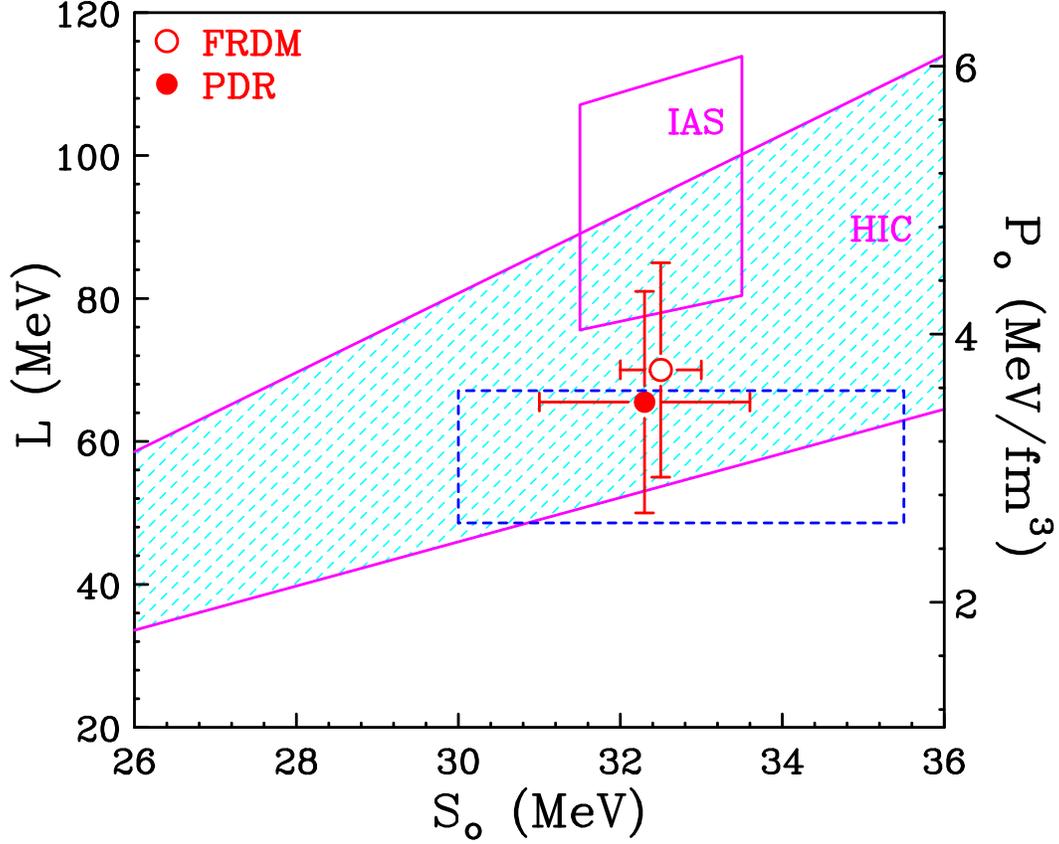}
\caption{(color online) Constraints on symmetry energy $\mathcal{S}_{\rm o}$ 
its first derivative $L$ and P$_{\rm o}$=P$_{\rm PNM}(\rho_{\rm o})$, all at 
saturation density, as derived from HIC \cite{tsang2009}, PDR
\cite{klim2007,carbone2010}, IAS \cite{dan2011} and FRDM \cite{moller2012}. 
Predictions of the consistent Skyrme parameterizations lie all within the blue
dashed rectangle. For full explanation see text and \protect\cite{tsang2012}.}
\label{fig:tsang}
\end{center}
\end{figure}

\begin{figure}[ht]
\begin{center}
\includegraphics[scale=0.6]{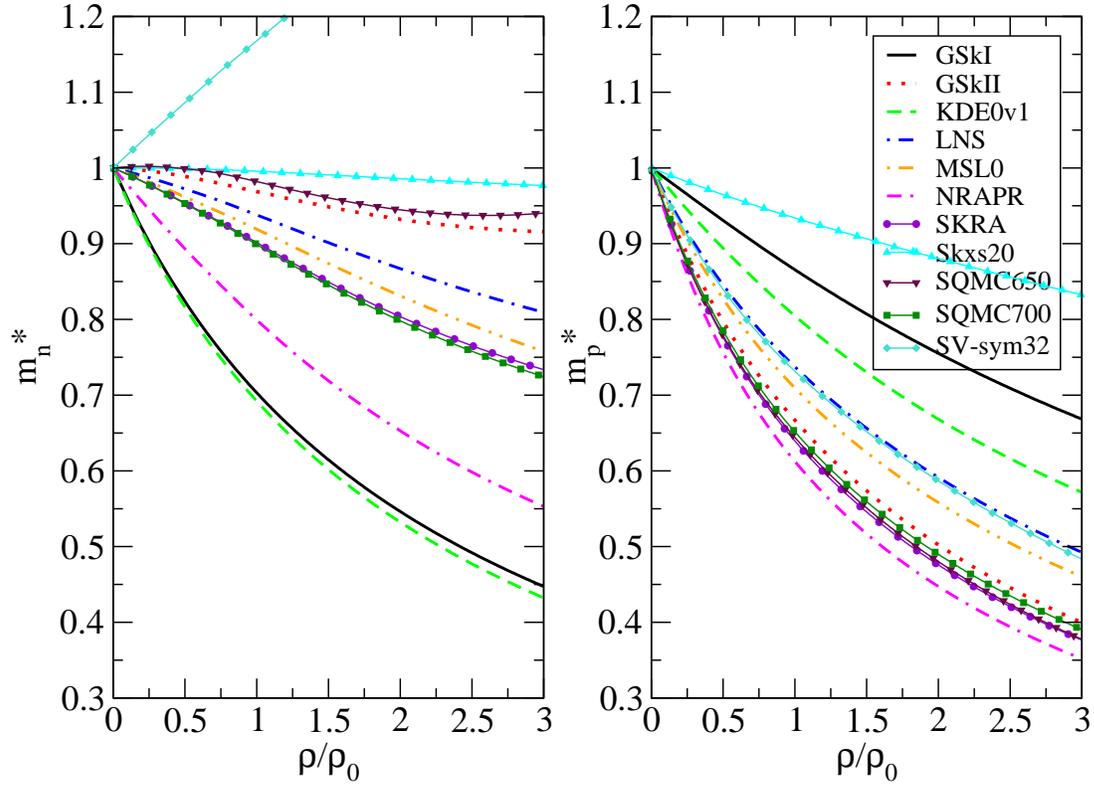}
\caption{(color online) Density dependence of neutron (left panel) and proton 
(right panel) effective mass in $\beta$-equilibrium matter as calculated by 
Skyrme interactions consistent with the macroscopic constraints. For more detail
see text.}
\label{fig:effmas}
\end{center}
\end{figure}

\begin{figure}[ht]
\begin{center}
\includegraphics[scale = 0.7]{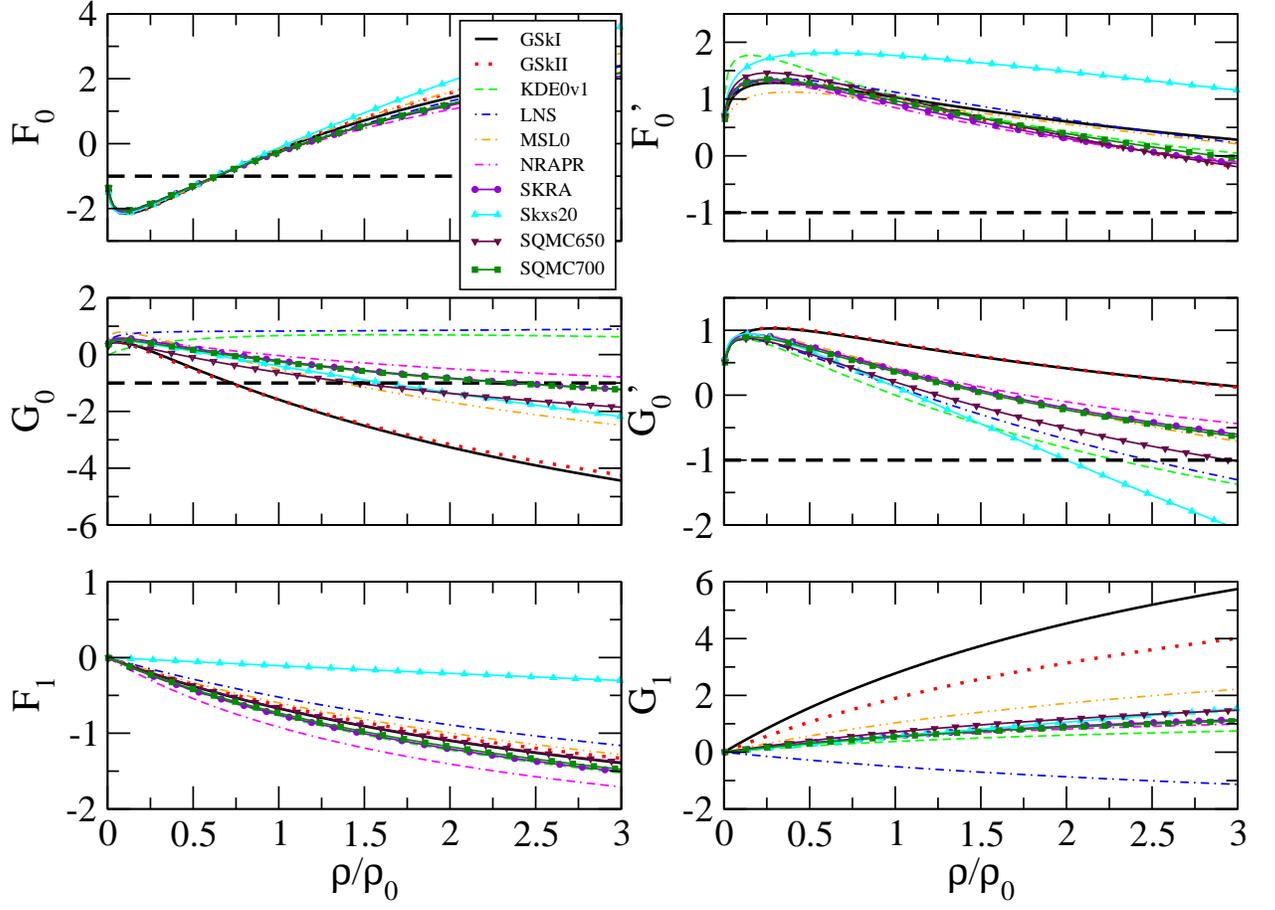}
\caption{(color online) Landau parameters calculated by CSkP sets, which passed 
microscopic constraints derived from the effective mass considerations, in SNM.
See text for more explanation.}
\label{fig:landau1}
\end{center}
\end{figure}

\begin{figure}[ht]
\begin{center}
\includegraphics[scale = 0.7]{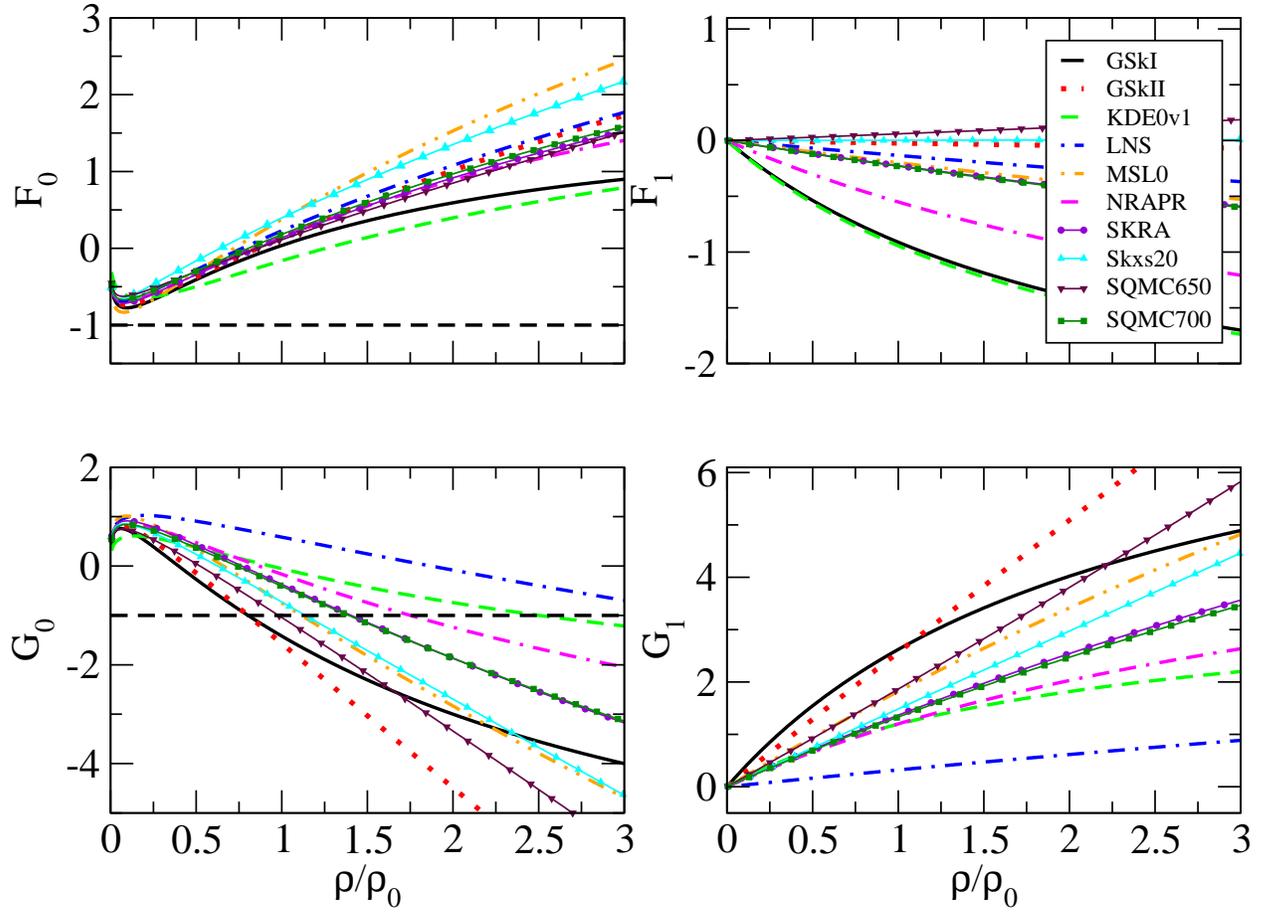}
\caption{(color online) The same as Fig.~\ref{fig:landau1}, but for PNM.}
\label{fig:landau2}
\end{center}
\end{figure}

\begin{figure}
\begin{center}
\includegraphics[scale=0.6]{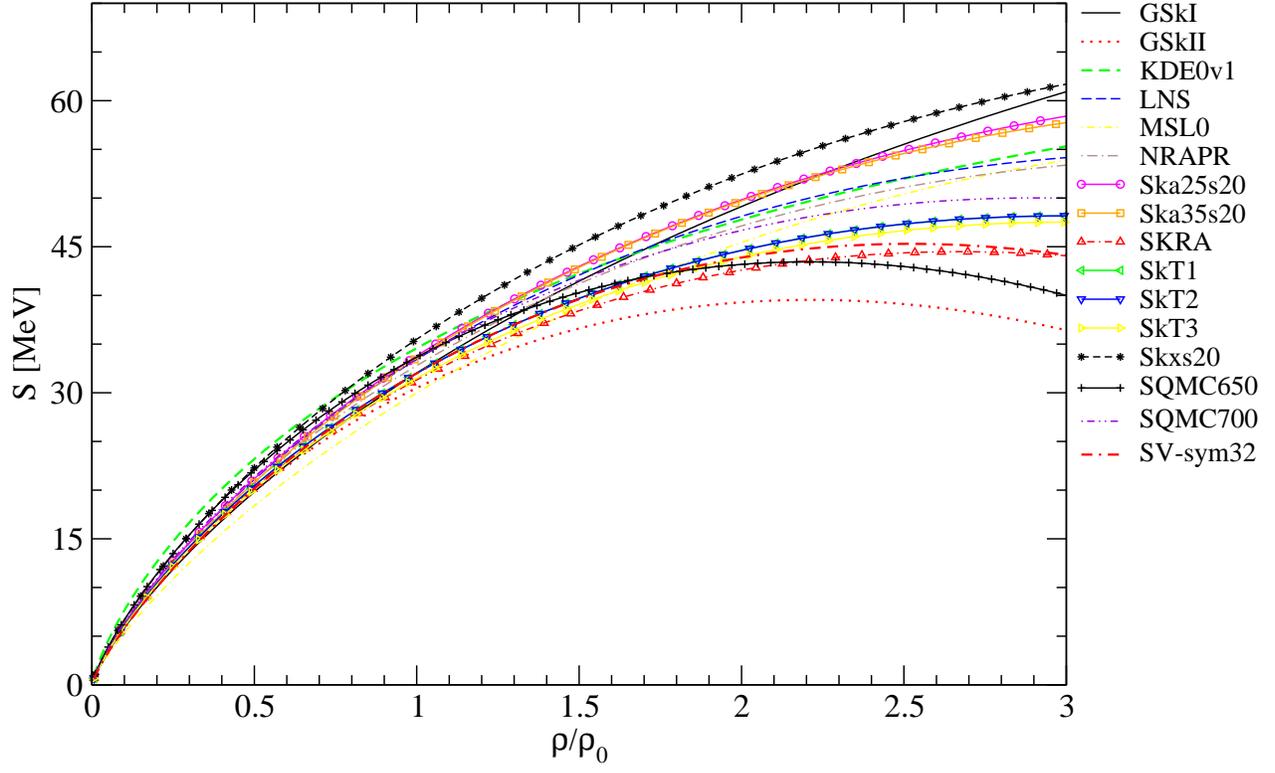}
\caption{(color online) Density dependence of the symmetry energy $\mathcal{S}$ 
as a function of $\frac{\rho}{\rho_{\rm o}}$ as calculated by Skyrme
interactions consistent with macroscopic constraints.}
\label{fig:esym}
\end{center}
\end{figure}

\begin{figure}
\begin{center}
\includegraphics[scale=0.6]{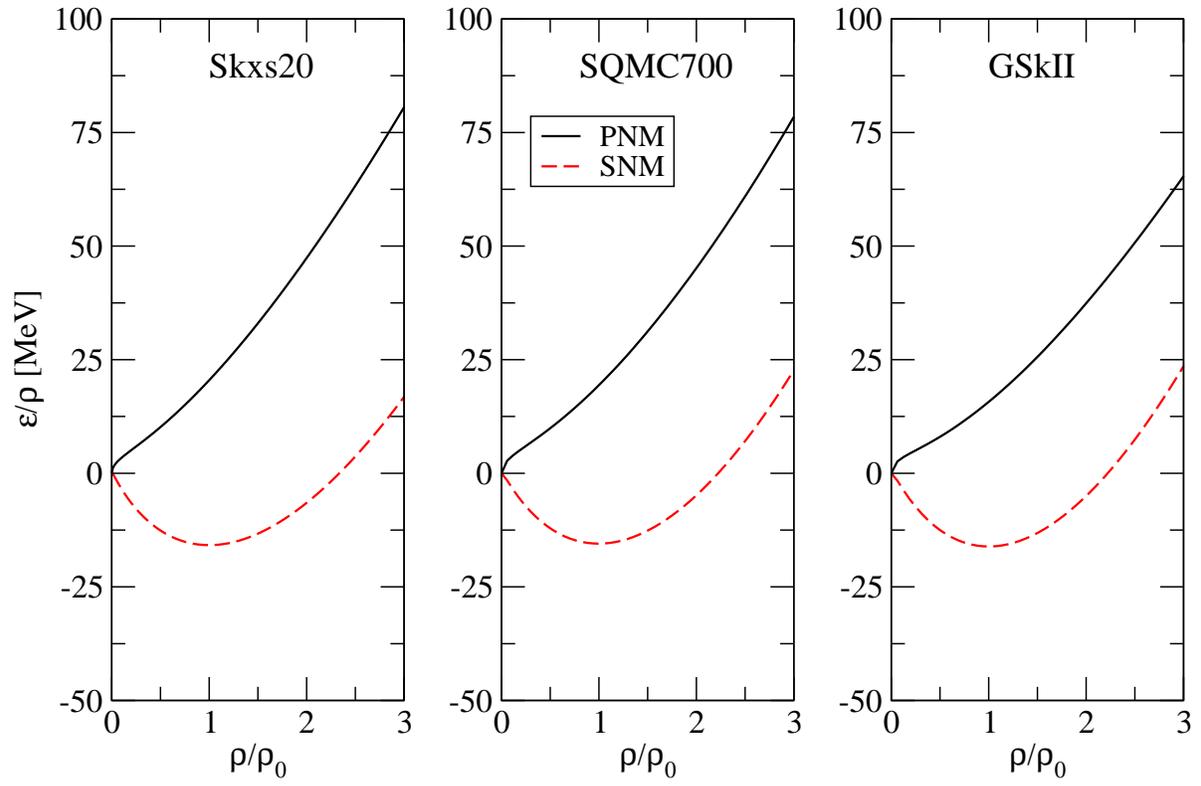}
\caption{(color online) Energy per particle in PNM and SNM  a function of 
particle number density $\rho$ for three selected Skyrme parameterizations
Skxs20, SQMC700 and GSkII.}
\label{fig:sym3f}
\end{center}
\end{figure}

\begin{figure}
\begin{center}
\includegraphics[scale=0.6]{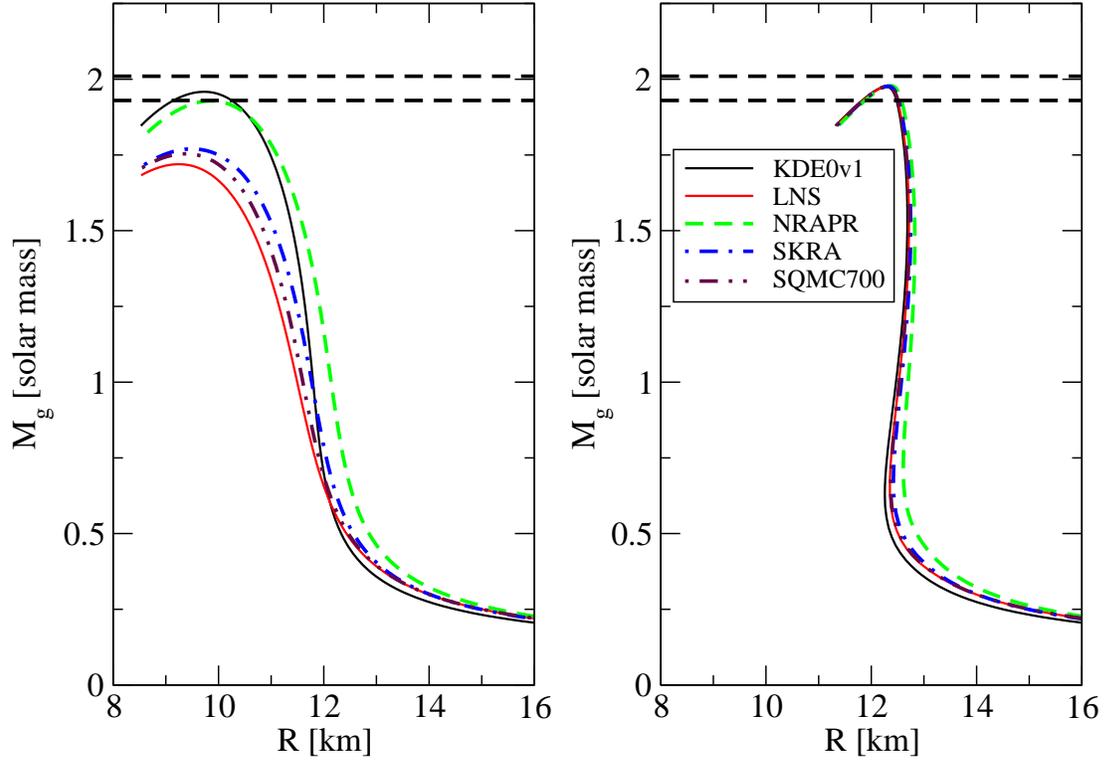}
\caption{(color online) Gravitational mass vs radius for cold non-rotational 
neutron stars as calculated using a Skyrme EoS augmented by BPS EoS at low
density \cite{baym1971} (left panel) and matched by a FQMC EoS at high densities
\cite{stone2007a} and by BPS EoS at low densities (right panel). The dashed 
lines indicate the limits on the maximum mass of the most massive neutron star
observed up-to-date \cite{demorest2010}.  Only Skyrme parameterizations which
are consistent with both macroscopic and microscopic constraints are used. For
more explanation see text.}
\label{fig:mr}
\end{center}
\end{figure}

\begin{figure}
\begin{center}
\includegraphics[scale=0.6]{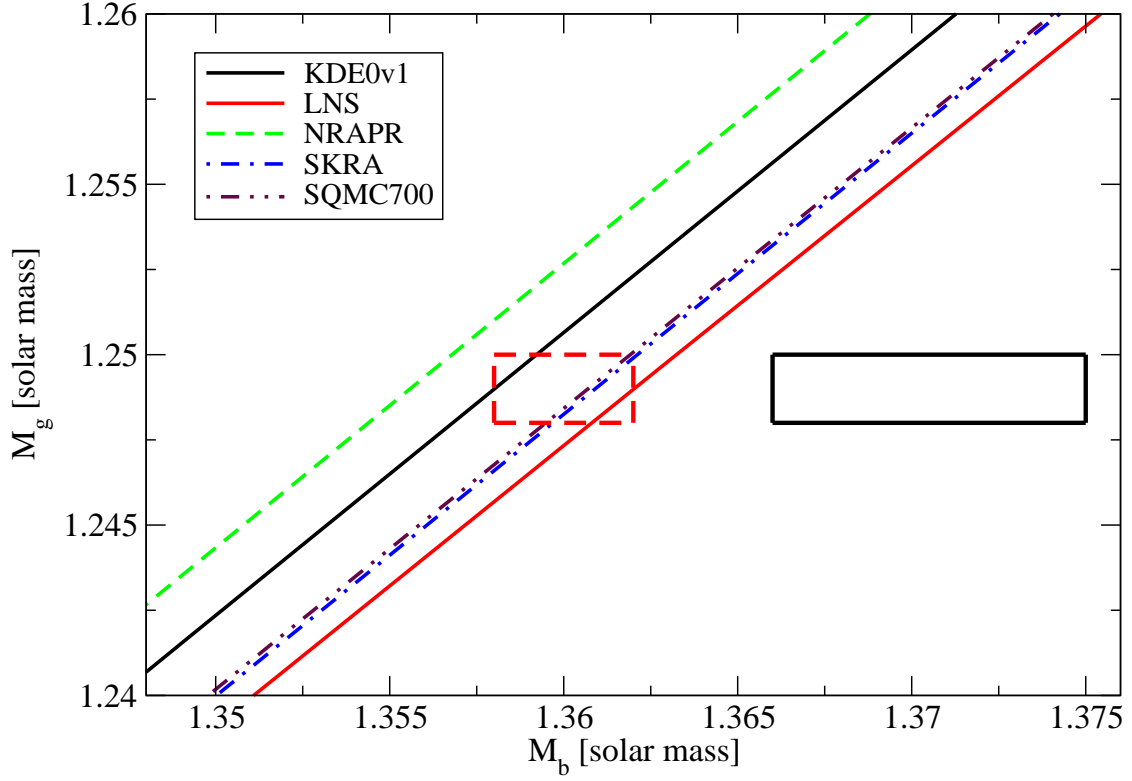}
\caption{(color online) Relation between the gravitational mass, M$_{\rm g}$, 
for the selected Skyrme models, and the corresponding baryonic mass, M$_{\rm
b}$. The boxes represent constraints derived by Podsiadlowski {\it et al.}
\cite{podsi2005} (full line box) and more recently by Kitaura {\it et al.}
\cite{kitaura2006} (dashed line box) based on the proposed properties of
system J0737-3039, as discussed in the text. Results are shown for Skyrme 
interaction consistent with both, macroscopic and microscopic constraints.}
\label{fig:pulb}
\end{center}
\end{figure}

\begin{figure}
\begin{center}
\includegraphics[scale=0.6]{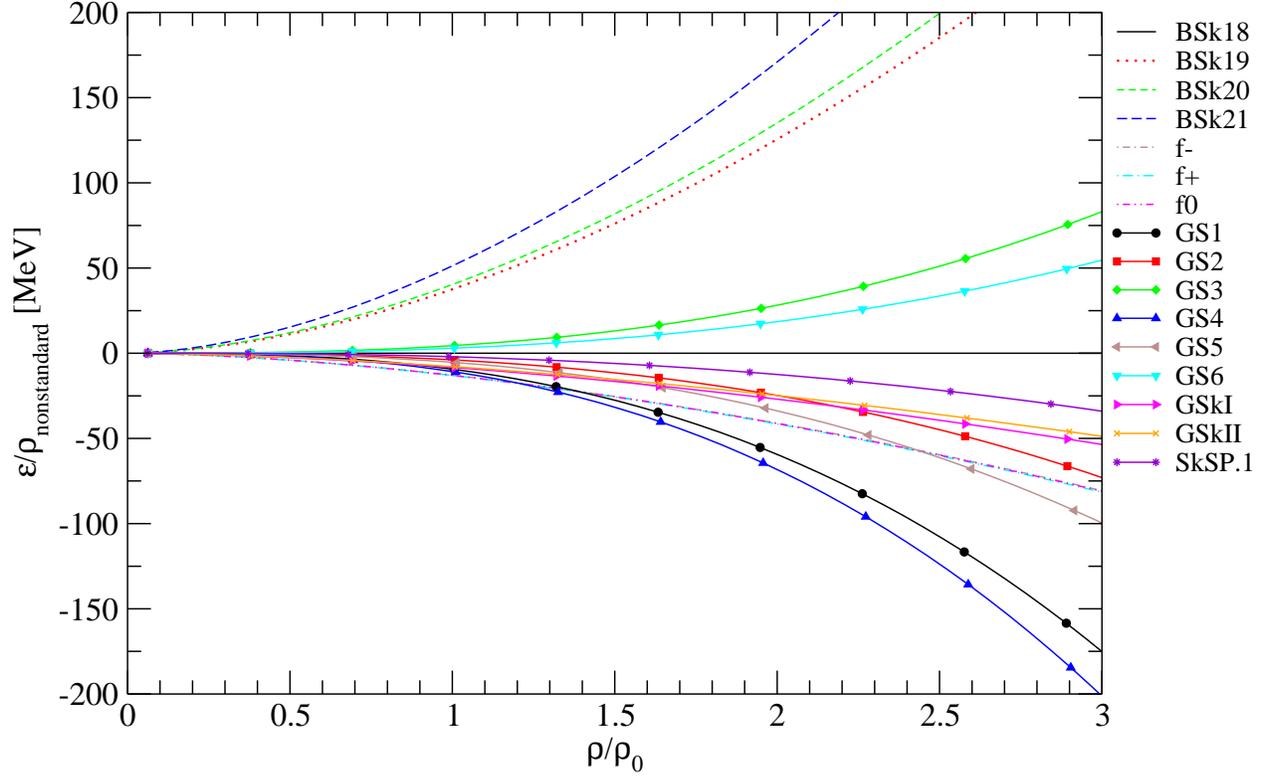}
\caption{(color online) Density dependence of the energy per particle, resulting
only from the contributions of the non-standard terms in
Eq.~(\ref{densityenergy}) for all non-standard Skyrme parameterizations used in
this work.}
\label{fig:nonstand}
\end{center}
\end{figure}

The authors thank Betty Tsang for providing Fig.~\ref{fig:tsang} prior to 
publication. JRS acknowledges helpful discussions with Anthony Thomas, Pierre
Guichon, Steven Moszkowski, P.-G. Reinhard, Bijay Agrawal, Shalom Shlomo and
Nick Stone.

This work was supported by the Brazilian agencies CNPq, CAPES and FAPESP. JRS
has the pleasure to thank for hospitality during her stay at Universidade
Federal Santa Catarina in the group of Debora Menezes, made possible by CNPq
Processo [401593/2009-6]. She also acknowledges with thanks the support under
United Kigdom STFC grant EP/D077133/1 and the stimulating environment at the
University of Surrey during the final stages of this work.  MD and OL thank for
hospitality in UFSC during initial stage of this work and Debora Menezes for
encouragement and  support.

\newpage
\begin{table}[hb!]
\scriptsize
% \squeezetable
\begin{ruledtabular}
\caption{List of  macroscopic constraints and the range of their 
experimental/empirical values, density region in which they are valid and the
corresponding range as found using successful Skyrme parameterizations (CSkP).
For more explanation see text.}
\centering
\begin{tabular}{lcccccc}
Constraint & Quantity      & Eq.  & Density Region       & Range of constraint   &
Range of constraint        & Ref.\\ 
      &                         &      &                      & exp/emp           &
 from CSkP         & \\ \hline
SM1   & K$_{\rm o}$             & (\ref{eq:incomp}),(\ref{eq:k0})  & $\rho_{\rm
o}$ (fm$^{-3}$) & 200 $-$ 260 MeV &  202.0 $-$ 240.3 MeV &
\cite{youngblood1999}\\
SM2   & K$^\prime=-$Q$_{\rm o}$ & (\ref{eq:skewness}),(\ref{eq:q0})  &
$\rho_{\rm o}$ (fm$^{-3}$) & 200 $-$ 1200 MeV&  362.5 $-$ 425.6 MeV &
\cite{farine1997}\\
SM3   & P($\rho$)               & (\ref{eq:pressnr})  & $2<\frac{\rho}{\rho_{\rm
o}}<3$ & Band Region   & see Fig.~\ref{fig:sm3}    & \cite{danielewicz2002}\\
SM4   & P($\rho$)               & (\ref{eq:pressnr}) &
$1.2<\frac{\rho}{\rho_{\rm o}}<2.2$    & Band Region  & see Fig.~\ref{fig:sm4}
  & \cite{lynch2009}\\
PNM1  & $\frac{E_{PNM}}{E_{PNM}^{\rm o}}$
                                & (\ref{eq:106}) & $0.014<\frac{\rho}{\rho_{\rm
o}}<0.106$   & Band Region &  see Fig.~\ref{fig:pnm1}
& \cite{schwenk2005,epelbaum2009}\\
PNM2  & P($\rho$)               & (\ref{eq:pressnr})  & $2<\frac{\rho}{\rho_{\rm
o}}<3$ & Band Region    & see Fig.~\ref{fig:pnm2}   & \cite{danielewicz2002}\\
\end{tabular}

\label{tab:listcon}
\end{ruledtabular}
\end{table}

\newpage
\begingroup
\centering
\squeezetable
\begin{longtable}{lcccrrrrrrrr}
\caption{Saturation properties of all Skyrme parameterization used in this work.
All entries are in MeV, except for the saturation density $\rho_{\rm o}$ in $fm^{-3}$ and the
dimensionless effective mass $m^* = M^*/M$.}
\label{tab:237} \\
\hline\hline
\multicolumn{1}{l}{Model} & \multicolumn{1}{c}{$\rho_{\rm o}$}
& \multicolumn{1}{c}{$E_{\rm o}$} & \multicolumn{1}{c}{$K_{\rm o}$}
& \multicolumn{1}{c}{$K'$} & \multicolumn{1}{c}{$J$}
& \multicolumn{1}{c}{$L$}  & \multicolumn{1}{c}{$K_{\rm sym}$} 
& \multicolumn{1}{c}{$Q_{\rm sym}$} & \multicolumn{1}{c}{$K_{\rm \tau,v}$}
& \multicolumn{1}{c}{$m^*$} \\
\hline 
\endfirsthead
\multicolumn{12}{c}%
{\tablename\ \thetable{} -- continued from previous page} \\
\hline\hline
\multicolumn{1}{l}{Model} & \multicolumn{1}{c}{$\rho_{\rm o}$}
& \multicolumn{1}{c}{$E_{\rm o}$} & \multicolumn{1}{c}{$K_{\rm o}$}
& \multicolumn{1}{c}{$K'$} & \multicolumn{1}{c}{$J$}
& \multicolumn{1}{c}{$L$}  & \multicolumn{1}{c}{$K_{\rm sym}$} 
& \multicolumn{1}{c}{$Q_{\rm sym}$} & \multicolumn{1}{c}{$K_{\rm \tau,v}$} 
& \multicolumn{1}{c}{$m^*$} \\
\hline 
\endhead
\hline
\multicolumn{12}{r}{{Continued on next page}} \\ 
\hline
\endfoot
\hline\hline
\endlastfoot
BSk1 \cite{samyn2002} & ~0.157~ & ~~-15.81~~ & ~~231.31~~ & 385.59 &
27.81 & 7.19 & ~~-281.83 & ~~606.46 & ~~-312.97 & ~~1.05 \\ 
BSk2 \cite{goriely2002} & 0.157 & -15.80 & 233.65 & 380.07 & 28.00 & 7.98 &
-296.98 & 557.91 & -331.87 & 1.04 \\ 
BSk2' \cite{goriely2002} & 0.157 & -15.79 & 233.32 & 380.86 & 28.00 & 7.79 &
-298.02 & 558.62 & -332.04 & 1.05 \\ 
BSk3 \cite{samyn2003} & 0.157 & -15.81 & 234.81 & 380.83 & 27.93 & 6.78 &
-306.90 & 550.34 & -336.59 & 1.12 \\ 
BSk4 \cite{goriely2003} & 0.157 & -15.77 & 236.84 & 367.17 & 28.00 & 12.54 &
-265.93 & 558.40 & -321.74 & 0.92 \\ 
BSk5 \cite{goriely2003} & 0.157 & -15.80 & 237.19 & 367.86 & 28.70 & 21.41 &
-240.30 & 499.92 & -335.56 & 0.92 \\ 
BSk6 \cite{goriely2003} & 0.157 & -15.75 & 229.14 & 370.64 & 28.00 & 16.84 &
-215.19 & 603.54 & -289.01 & 0.80 \\ 
BSk7 \cite{goriely2003} & 0.157 & -15.76 & 229.26 & 370.92 & 28.00 & 17.99 &
-209.35 & 598.16 & -288.18 & 0.80 \\
\end{longtable}
\pagebreak
\endgroup

\newpage
\begin{landscape}
\begingroup
\centering
\squeezetable
\begin{longtable}{lccccccccccclccccccccccc}
\caption{Constraints for the Skyrme Models. The $^{\dag}$ symbol on the minus
sign indicates that the model satisfies all the constraints except one.}
\label{tab:const} \\
\hline\hline
\multicolumn{1}{l}{Model}   & \multicolumn{1}{c}{SM1}
& \multicolumn{1}{c}{SM2}   & \multicolumn{1}{c}{SM3}
& \multicolumn{1}{c}{SM4}   & \multicolumn{1}{c}{PNM1}
& \multicolumn{1}{c}{PNM2}  & \multicolumn{1}{c}{MIX1}
& \multicolumn{1}{c}{MIX2}  & \multicolumn{1}{c}{MIX3} 
& \multicolumn{1}{c}{MIX4}  & \multicolumn{1}{c}{MIX5}
& \multicolumn{1}{l}{Model} & \multicolumn{1}{c}{SM1}
& \multicolumn{1}{c}{SM2}   & \multicolumn{1}{c}{SM3}
& \multicolumn{1}{c}{SM4}   & \multicolumn{1}{c}{PNM1}
& \multicolumn{1}{c}{PNM2}  & \multicolumn{1}{c}{MIX1}
& \multicolumn{1}{c}{MIX2}  & \multicolumn{1}{c}{MIX3}
& \multicolumn{1}{c}{MIX4}  & \multicolumn{1}{c}{MIX5} \\
\hline 
\endfirsthead
\multicolumn{24}{c}%
{\tablename\ \thetable{} -- continued from previous page} \\
\hline\hline
\multicolumn{1}{l}{Model}   & \multicolumn{1}{c}{SM1}
& \multicolumn{1}{c}{SM2}   & \multicolumn{1}{c}{SM3}
& \multicolumn{1}{c}{SM4}   & \multicolumn{1}{c}{PNM1}
& \multicolumn{1}{c}{PNM2}  & \multicolumn{1}{c}{MIX1}
& \multicolumn{1}{c}{MIX2}  & \multicolumn{1}{c}{MIX3} 
& \multicolumn{1}{c}{MIX4}  & \multicolumn{1}{c}{MIX5}
& \multicolumn{1}{l}{Model} & \multicolumn{1}{c}{SM1}
& \multicolumn{1}{c}{SM2}   & \multicolumn{1}{c}{SM3}
& \multicolumn{1}{c}{SM4}   & \multicolumn{1}{c}{PNM1}
& \multicolumn{1}{c}{PNM2}  & \multicolumn{1}{c}{MIX1}
& \multicolumn{1}{c}{MIX2}  & \multicolumn{1}{c}{MIX3}
& \multicolumn{1}{c}{MIX4}  & \multicolumn{1}{c}{MIX5} \\
\hline 
\endhead
\hline
\multicolumn{24}{l}{{Continued on next page}} \\ 
\hline
\endfoot
\hline\hline
\endlastfoot
BSk1 & + & + & + & + & + & - & - & - & - & + & -  & GS1 & + & + & - & + & + & +
& - & + & - & + & +  \\ 
BSk2 & + & + & + & + & + & - & - & - & - & + & -  & GS2 & - & + & - & - & + & -
& - & - & - & + & -  \\ 
BSk2' & + & + & + & + & + & - & - & - & - & + & -  & GS3 & - & - & - &  -  & + &
- & - & - & + & + & -  \\ 
BSk3 & + & + & + & + & + & - & - & - & - & + & -  & GS4 & + & + & - & - & + & -
& - & - & - & - & +  \\ 
BSk4 & + & + & + & + & + & - & - & - & - & + & -  & GS5 & - & + & - & - & + & -
& - & - & - & - & +  \\ 
BSk5 & + & + & + & + & + & - & - & - & - & + & -  & GS6 & - & - & - & - & + & -
& - & - & - & - & -  \\ 
BSk6 & + & + & + & + & + & - & - & - & - & + & +  & GSkI & + & + & + & + & + & +
& + & + & -$^{\dag}$ & + & +  \\ 
BSk7 & + & + & + & + & + & - & - & - & - & + & +  & GSkII & + & + & + & + & + &
+ & + & + & -$^{\dag}$ & + & +  \\
\end{longtable}
\pagebreak
\endgroup
\end{landscape}

\newpage
\begingroup
\centering
\squeezetable
\begin{longtable}{lrrccccrrrrr}
\caption{List of the Skyrme parameterizations and their compliance with
macroscopic constraints used in this work. For SM1, SM2 and MIX1-5 we give the
standard deviation; the constraint is satisfied if the standard deviation is
less of equal one. For SM3, SM4, PNM1 and PNM2 we give the fraction of full
density range in \% in which the constraint is not satisfied. Letter L (H)
indicates that the Skyrme prediction starts to fail in the beginning (end) of
the density range. No indication of letters means that the model fails in the
middle of the density range. For models that fail in $100\%$, the letter U (D)
indicates that its curve is above (below) the band defined by the constraint.
The $^\dag$ symbol marks a constraint which is the only one which is not
satisfied (out of 11) by a particular Skyrme parameterization. For more details
see text.}
\label{tab:const-num} \\
\hline\hline
\multicolumn{1}{l}{Model}         & \multicolumn{1}{r}{SM1}
& \multicolumn{1}{r}{SM2}         & \multicolumn{1}{c}{SM3($\%$)}
& \multicolumn{1}{c}{SM4($\%$)}   & \multicolumn{1}{c}{PNM1($\%$)}
& \multicolumn{1}{c}{PNM2($\%$)}  & \multicolumn{1}{r}{MIX1} 
& \multicolumn{1}{r}{MIX2}        & \multicolumn{1}{r}{MIX3} 
& \multicolumn{1}{r}{MIX4}        & \multicolumn{1}{r}{MIX5} \\
\hline 
\endfirsthead
\multicolumn{12}{c}%
{\tablename\ \thetable{} -- continued from previous page} \\
\hline\hline
\multicolumn{1}{l}{Model}         & \multicolumn{1}{r}{SM1}
& \multicolumn{1}{r}{SM2}         & \multicolumn{1}{c}{SM3($\%$)}
& \multicolumn{1}{c}{SM4($\%$)}   & \multicolumn{1}{c}{PNM1($\%$)}
& \multicolumn{1}{c}{PNM2($\%$)}  & \multicolumn{1}{r}{MIX1} 
& \multicolumn{1}{r}{MIX2}        & \multicolumn{1}{r}{MIX3} 
& \multicolumn{1}{r}{MIX4}        & \multicolumn{1}{r}{MIX5} \\
\hline 
\endhead
\hline
\multicolumn{12}{r}{{Continued on next page}} \\ 
\hline
\endfoot
\hline\hline
\endlastfoot
BSk1 & 0.04 & -0.63 & 0 & 0 & 0 &  100D & -1.88 & -2.82 & 1.30 & 0.69 & 1.70 \\ 
BSk2 & 0.12 & -0.64 & 0 & 0 & 0 &  100D & -1.80 & -2.78 & 1.21 & 0.69 & 3.80 \\ 
BSk2' & 0.11 & -0.64 & 0 & 0 & 0 &  100D & -1.80 & -2.79 & 1.21 & 0.69 & 3.90
\\ 
BSk3 & 0.16 & -0.64 & 0 & 0 & 0 &  100D & -1.83 & -2.85 & 1.18 & 0.69 & 4.40 \\ 
BSk4 & 0.23 & -0.67 &  2.0L & 0 & 0 &  100D & -1.80 & -2.53 & 1.26 & 0.54 & 2.00
\\ 
BSk5 & 0.24 & -0.66 &  2.7L & 0 & 0 &  100D & -1.52 & -2.03 & 1.19 & 0.31 & 1.20
\\ 
BSk6 & -0.03 & -0.66 & 0 & 0 & 0 &  100D & -1.80 & -2.29 & 1.43 & 0.54 & 0.90
\\ 
BSk7 & -0.02 & -0.66 & 0 & 0 & 0 &  100D & -1.80 & -2.22 & 1.43 & 0.54 & 0.80
\\
\end{longtable}
\pagebreak
\endgroup
\newpage
\begin{table}
\begin{ruledtabular}
\caption{Number of Skyrme models consistent with individual macroscopic
constraints.}
\centering
\begin{tabular}{lccccccccccc}
Constraints  & SM1   & SM2   & SM3   & SM4   & PNM1  & PNM2 & MIX1  & MIX2   &
MIX3  & MIX4  & MIX5 \\
No CSkP      & $202$ & $214$ & $145$ & $211$ & $186$ & $94$ & $139$ & $102$ 
& $65$ & $198$ & $159$ \\
\end{tabular}
\label{tab:nap}
\end{ruledtabular}
\end{table}
\begin{table}[!htb]
\scriptsize
% \squeezetable
\begin{ruledtabular}
\caption{Parameters of the Skyrme interactions consistent with the macroscopic
constraints. $t_0$ in MeV.fm$^3$,  $t_1$, $t_2$ in
MeV.fm$^5$; $t_{3i}$ in MeV.fm$^{3+3\sigma_i}$. $x_0$, $x_1$, $x_2$, $x_{3i}$,
and $\sigma_i$ are dimensionless. For all parameterizations
$t_4=x_4=0$ and $t_5=x_5=0$.}
\centering
\begin{tabular}{lccccccccccccccc}
Skyrme & $t_0$ & $t_1$ & $t_2$ & $t_{31}$ & $t_{32}$ & $t_{33}$ & $x_0$ & $x_1$ 
& $x_2$ & $x_{31}$ & $x_{32}$ & $x_{33}$ & $\sigma_1$ & $\sigma_2$ & $\sigma_3$
\\ \hline
GSkI & -1855.5 & 397.2 & 264.6 & 13858.0 & -2694.1 & -319.9 & 0.12 & -1.76 &
-1.81 & 0.13 & -1.19 & -0.46 & 0.33 & 0.67 & 1.00 \\ 
GSkII    & -1856.0 & 393.1 & 266.1  & 13842.9 & -2689.7 & - & 0.09 & -0.72 &
-1.84 & -0.10 & -0.35 & - & 0.33 & 0.67 & -  \\ 
KDE0v1   & -2553.1 & 411.7 & -419.9 & 14603.6 & - & - & 0.65  & -0.35 &-0.93 &
0.95 & - & - & 0.17 & - & -   \\ 
LNS      & -2485.0 & 266.7 & -337.1 & 14588.2 & - & - & 0.06  & 0.66  &-0.95
&-0.03 & - & - & 0.17 & - & - \\ 
MSL0     & -2118.1 & 395.2 & -64.0  & 12875.7 & - & - &-0.07  & -0.33 & 1.36
&-0.23 & - & - & 0.24 & - & - \\ 
NRAPR    & -2719.7 & 417.6 & -66.7  & 15042.0 & - & - & 0.16  & -0.05 & 0.03 &
0.14 & - & - & 0.14 & - & -  \\ 
Ska25s20 & -2180.5 & 281.5 & -160.4 & 14577.8 & - & - & 0.14  & -0.80 & 0.00 &
0.06 & - & - & 0.25 & - & - \\
\end{tabular}
\label{tab:par}
\end{ruledtabular}
\end{table}
\newpage
\vspace{1.2cm}
\begin{table}[!htb]
\scriptsize
% \squeezetable
\begin{ruledtabular}
\caption{Properties of nuclear matter at saturation as calculated using the
Skyrme parameterizations consistent with the macroscopic constraints. All
entries are in MeV except for dimensionless $\frac{\mathcal{S}(\rho_{\rm
o}/2)}{J}$, $\frac{3P_{\rm PNM}}{L \rho_{\rm o}}$ and $m^*$.}
\centering
\begin{tabular}{lccccccccccc}
Model & $\rho_{\rm o}$ & $E_{\rm o}$ & $K_{\rm o}$ & $K^{\prime}$ & $J$ & $L$  &
$K_{\rm vs}$ & $\frac{\mathcal{S}(\rho_{\rm o}/2)}{J}$ & $\frac{3P_{\rm PNM}}{L
\rho_{\rm o}}$ &  $m^*$ \\ \hline
GSkI & 0.159 & -16.02 & 230.21 & 405.58 & 32.03 & 63.45 & -364.19 & 0.62 & 1.02
& 0.78 \\ 
GSkII & 0.159 & -16.12 & 233.40 & 398.73 & 30.49 & 48.63 & -366.54 & 0.65 & 1.09
& 0.79 \\ 
KDE0v1 & 0.165 & -16.23 & 227.54 & 384.86 & 34.58 & 54.69 & -362.78 & 0.67 &
1.03 & 0.74 \\ 
LNS & 0.175 & -15.32 & 210.78 & 382.55 & 33.43 & 61.45 & -384.55 & 0.63 & 1.06 &
0.83 \\ 
MSL0 & 0.160 & -16.00 & 230.00 & 380.32 & 30.00 & 60.00 & -360.11 & 0.61 & 1.06
& 0.80 \\ 
NRAPR & 0.161 & -15.85 & 225.65 & 362.54 & 32.78 & 59.63 & -385.32 & 0.63 & 1.08
& 0.69 \\ 
Ska25s20 & 0.161 & -16.07 & 220.75 & 413.45 & 33.78 & 63.81 & -381.56 & 0.63 &
1.02 & 0.98 \\ 
Ska35s20 & 0.158 & -16.08 & 240.27 & 378.65 & 33.57 & 64.83 & -407.11 & 0.62 &
1.02 & 1.00 \\ 
SKRA     & 0.159 & -15.78 & 216.98 & 378.76 & 31.32 & 53.04 & -364.92 & 0.65 &
1.09 & 0.75 \\
SkT1 & 0.161 & -15.98 & 236.16 & 383.52 & 32.02 & 56.18 & -380.68 & 0.64 & 1.03
& 1.00 \\ 
SkT2 & 0.161 & -15.94 & 235.73 & 382.67 & 32.00 & 56.16 & -380.48 & 0.64 & 1.03
& 1.00 \\ 
SkT3 & 0.161 & -15.95 & 235.74 & 382.70 & 31.50 & 55.31 & -374.14 & 0.64 & 1.03
& 1.00 \\ 
Skxs20 & 0.162 & -15.81 & 201.95 & 425.56 & 35.50 & 67.06 & -383.37 & 0.63 &
1.03 & 0.96 \\
SQMC650 & 0.172 & -15.57 & 218.11 & 376.75 & 33.65 & 52.92 & -399.28 & 0.66 &
1.10 & 0.78 \\ 
SQMC700 & 0.171 & -15.49 & 222.20 & 369.94 & 33.47 & 59.06 & -396.85 & 0.64 &
1.09 & 0.76 \\ 
SV-sym32 & 0.159 & -15.94 & 233.81 & 380.11 & 32.00 & 57.07 & -398.44 & 0.63 &
1.08 & 0.90 \\ 
\end{tabular}
\label{tab:18}
\end{ruledtabular}
\end{table}
\newpage
\begin{table}[!htb] 
\scriptsize
% \squeezetable
\begin{ruledtabular}
\caption{Selected properties on nuclear matter as predicted by consistent Skyrme
parameterizations at 
$3\rho_{\rm o}$. $L(\rho_{\rm o})$ is included for a comparison. All
quantities are in MeV except for $m^*$, which is dimensionless.}
\centering
\begin{tabular}{lcccrrr} 
Force & $m^*$ & $\mathcal{S}$ & $L(\rho_{\rm o})$ & L(3$\rho_{\rm o})$ & 
$K_{\rm sym}$  & $Q_{\rm sym}$\\\hline
GSkI     & 0.536 & 60.92 &  63.45  & 89.71   & -253.92  & 947.56 \\ 
GSkII    & 0.557 & 36.42 &  48.63  & -69.49  & -731.84  & 1226.76\\ 
KDE0v1   & 0.493 & 55.31 &  54.69  & 53.66   & -192.75  & 1366.50\\ 
LNS      & 0.612 & 54.15 &  61.45  & 23.06   & -515.88  & 1027.81\\ 
MSL0     & 0.571 & 53.91 &  60.00  & 50.80   & -424.29  & 776.89 \\
NRAPR    & 0.430 & 53.38 &  59.63  & 27.72   & -468.17  & 1024.00\\ 
Ska25s20 & 0.942 & 58.41 &  63.81  & 51.87   & -411.53  & 1047.02\\ 
Ska35s20 & 1.000 & 57.75 &  64.83  & 43.51   & -473.55  & 1027.22\\ 
SKRA     & 0.748 & 44.07 &  53.04  & -22.54  & -596.94  & 1093.69\\
SkT1     & 1.000 & 48.17 &  56.18  & -0.31   & -527.96  & 1158.54\\ 
SkT2     & 1.000 & 48.17 &  56.16  & -0.14   & -527.25  & 1157.11\\ 
SkT3     & 1.000 & 47.53 &  55.31  & 0.78    & -515.94  & 1134.50\\ 
Skxs20   & 0.899 & 61.70 &  67.06  & 56.83   & -428.23  & 1058.16\\
SQMC650  & 0.779 & 33.65 &  52.92  & -76.28  & -173.15  & 349.74 \\
SQMC700  & 0.507 & 49.98 &  59.06  & -6.52   & -603.34  & 1109.31\\ 
SV-sym32 & 0.749 & 44.14 &  57.07  & -42.43  & -759.99  & 1060.52\\ 
\end{tabular}
\label{tab:3rho}
\end{ruledtabular}
\end{table}

\newpage
\begin{table}[!htb] 
\scriptsize
% \squeezetable
\begin{ruledtabular}
\caption{Relevant combinations of Skyrme parameters of consistent Skyrme
parameterizations: $C_{\rm o}^{\rho^\prime}$, and $C_1^{\rho^\prime}$ are given
in MeV.fm$^3$; $C_{\rm o}^{\tau^\prime}$, and $C_1^{\tau^\prime}$ in
MeV.fm$^5$;  $\Theta_{\rm s}^\prime$, $\Theta_{\rm v}^\prime$, $\Theta_{\rm
sym}^\prime$ and $\Theta_{\rm n}^\prime$ in MeV.fm$^5$ \cite{chabanat1997}. For
full explanation, see text.}
\centering
\begin{tabular}{lcrrrrrrr}
Force    &  $C_{\rm o}^{\rho^\prime}$ & $C_1^{\rho^\prime}$ & 
$C_{\rm o}^{\tau^\prime}$ & $C_1^{\tau^\prime}$ &  $\Theta_{\rm s}^\prime$  & 
$\Theta_{\rm v}^\prime$ & $\Theta_{\rm sym}^\prime$ & $\Theta_{\rm n}^\prime$
\\ \hline
GSkI   &  -279.2 & 68.2  & 37.6 &  19.3 & 602.3  &  147.2 &  -763.6 &  455.28\\
GSkII  &  -276.9 & 154.3 & 34.7 & -33.6 & 554.6  &  546.4 &   530.1 &   8.17 \\
KDE0v1 &  -282.5 & 81.6  & 43.3 &  14.5 & 692.3  &  229.8 &  -695.0 &  462.4 \\
LNS    &  -250.3 & 137.9 & 25.0 & -19.5 & 400.8  &  356.4 &   267.6 &   44.4 \\
MSL0   &  -272.7 & 132.7 & 32.4 & -23.1 & 518.4  &  444.3 &   296.2 &   74.0 \\
NRAPR  &  -297.7 & 143.5 & 57.0 & -28.0 & 912.2  &  680.1 &   215.7 &  232.2 \\
Ska25s20 &-240.9 & 131.2 &  2.6 &   0.5 &  42.3  &   16.9 &   -33.8 &   25.4 \\
Ska35s20 &-240.5 & 135.4 &  0.0 &   0.0 & -0.12  &  -0.04 &    -0.1 &  -0.07 \\
SKRA    & -283.8 & 152.5 & 43.7 & -33.1 & 699.7  &  615.0 &   445.5 &   84.7 \\
SkT1    & -237.1 & 122.3 &  0.0 &   0.0 &   0.0  &    0.0 &     0.0 &    0.0 \\
SkT2    & -236.9 & 122.1 &  0.0 &   0.0 &   0.0  &    0.0 &     0.0 &    0.0 \\
SkT3    & -236.9 & 119.0 &  0.0 &   0.0 &   0.0  &    0.0 &     0.0 &    0.0 \\
Skxs20  & -240.6 & 148.9 &  4.8 &  -4.9 &  76.9  &   77.7 &    79.4 &   -0.8 \\
SQMC650 & -263.8 & 168.0 & 34.3 & -36.8 & 548.6  &  568.3 &   607.8 &  -19.7 \\
SQMC700 & -269.6 & 150.6 & 39.3 & -29.2 & 629.5  &  548.6 &   386.8 &   80.9 \\
SV-sym32& -253.7 & 181.6 & 14.5 & -37.4 & 232.0  &  415.3 &   781.8 & -183.3 \\
\end{tabular}
\label{tab:comb}
\end{ruledtabular}
\end{table}

\newpage
\begin{table}[!htb] 
% \scriptsize
\tiny
\squeezetable
\begin{ruledtabular}
\caption{Brief compilation of methods used in fitting of the Skyrme interactions
consistent with the macroscopic constraints  and main data used in the fit. For
full explanation and details, see original papers.}
\centering
\begin{tabular}{lll}
Force    &    Method &  Data used for fit \\ \hline
GSkI,GSkII &  simulated annealing & ground state properties of normal and exotic
              nuclei:\\
           &  method              & binding energies, charge radii, \\
           &                      & radii for 1d$_{\rm 5/2}$and 1f$_{\rm 7/2}$
                                    neutron orbits in $^{\rm 17}$O and
                                    $^{\rm 41}$Ca, \\
         &                        & breathing mode energies for $^{\rm 90}$Zr
                                   and $^{\rm 208}$Pb,\\
         &                        & single-particle energies in $^{\rm
                                    208}$Pb, \\
         &                        & EoS of pure neutron matter
\cite{wiringa1988} \\ \hline        
KDE0v1   &  simulated annealing   & ground state properties of normal and exotic
nuclei:\\
\end{tabular}
\label{tab:fit}
\end{ruledtabular}
\end{table}

\newpage
\vspace{0.8cm}

\vspace{0.5cm}

\newpage
\vspace*{1.2cm}

\newpage
\vspace*{1.2cm}

\newpage
\vspace*{1.2cm}

\newpage
\vspace{1cm}

\newpage
\vspace{0.8cm}

\newpage
\vspace{0.8cm}

\newpage
\vspace{0.8cm}

\newpage

\vspace{0.8cm}
  
\newpage
\vspace{1cm}

\vspace{1cm}

\vspace{1cm}

\end{document}